\documentclass{pasj00}

\begin{document}
\SetRunningHead{Hirashita et al.}{AKARI
Observation of Blue Compact Dwarf Galaxies}
\Received{2008/05/06}
\Accepted{2008/08/07}

\title{Far-Infrared Properties of Blue Compact
Dwarf Galaxies Observed with AKARI/Far-Infrared
Surveyor (FIS)\thanks{Based on observations with AKARI, a
JAXA project with the participation of ESA.}}

\author{Hiroyuki \textsc{Hirashita}}
\affil{Institute of Astronomy and Astrophysics,
 Academia Sinica, P.O. Box 23-141, Taipei 106, Taiwan}
\email{hirashita@asiaa.sinica.edu.tw}
\author{Hidehiro \textsc{Kaneda}}
\affil{Institute of Space and Astronautical Science,
   Japan Aerospace Exploration Agency, 3-1-1, Yoshinodai,
   Sagamihara, Kanagawa 229-8510}
\author{Takashi \textsc{Onaka}}
\affil{Department of Astronomy, Graduate School of
   Science, The University of Tokyo, 7-3-1 Hongo, Bunkyo-ku,
   Tokyo 113-0033}
\and
\author{Toyoaki \textsc{Suzuki}}
\affil{Advanced Technology Center, National Astronomical Observatory
of Japan, 2-21-1 Osawa, Mitaka, Tokyo 181-8588}


\maketitle

\begin{abstract}
We report basic far-infrared (FIR) properties of eight blue
compact dwarf galaxies (BCDs) observed by AKARI.
We measure the fluxes at the four FIS bands
(wavelengths of 65 $\mu$m, 90 $\mu$m, 140 $\mu$m, and
160 $\mu$m). Based on these fluxes, we estimate basic
quantities about dust: dust temperature, dust mass,
and total FIR luminosity. We find that the typical dust
temperature of the BCD sample is systematically higher than
that of normal spiral galaxies, although there is a
large variety. The
interstellar radiation field estimated from the dust
temperature ranges up to 100 times of the Galactic value.
This confirms the concentrated star-forming activity in
BCDs. The star formation rate can be evaluated from the FIR luminosity
as 0.01--0.5 $M_\odot$ yr$^{-1}$. Combining this quantity with
gas mass taken from the literature, we estimate the gas consumption
timescales (gas mass divided by the star formation rate), which prove to
span a wide range from
1 Gyr to 100 Gyr. A natural
interpretation of this large variety can be provided by intermittent
star formation activity. We finally show the relation between
dust-to-gas ratio
and metallicity (we utilize our estimate of dust mass, and take
other necessary quantities from the literature). There is a
positive correlation between dust-to-gas ratio and metallicity
as expected from chemical
evolution models.
\end{abstract}

\section{Introduction}\label{sec:intro}

Dust grains absorb stellar ultraviolet (UV)--optical light
and reprocess it into far-infrared (FIR), thereby affecting
the energetics of interstellar medium. In particular,
\citet{hirashita_ferrara02} show that the
dust produced by supernovae efficiently absorbs
UV heating photons and helps gas cooling and H$_2$
production even in primeval galaxies
(typical age is $<10^8$ yr). Dust grains
also affect the escape fraction of UV photons out of
galaxies. Observationally, this means that there is
always a correction factor for dust extinction when
one estimates star formation rate (SFR) from UV luminosity
\citep{buat96}. Indeed, a large correction
factor for dust extinction is required for the cosmic star
formation rate traced with UV
\citep{steidel99,takeuchi05,lefloch05,perez05}.
Also theoretically, how dust accumulates in the history
of the Universe is important in the context of
the cosmic reionization, because dust absorption
affects the escape fraction of ionizing photons
\citep{ciardi02}.

The origin of the cosmic dust is important in itself, and
the formation and destruction of grains are tightly related
to the star formation activity of galaxies. The condensation
of heavy elements is an important process for
the formation of grains. In particular, in young galaxies,
dust is predominantly condensed in Type II supernovae
(SNe II), whose progenitors have short lifetimes
\citep{dwek80,kozasa89,morgan02,nozawa03}.
However, how much dust forms in a SN II is not known.
There is a large difference among observationally derived
dust mass
\citep{moseley89,dunne03,hines04,sugerman06,meikle07,
ercolano07,rho08}.
Thus, it is important to constrain the dust mass formed in
SNe II by quantifying dust mass in young galaxies.

A large amount of dust has been suggested to exist at high
redshift ($z$) (e.g., \cite{bertoldi03,priddey03}). The
estimate of dust mass in high-$z$ galaxies indeed
constrains how much dust forms in SNe II \citep{dwek07}.
The rest-frame UV extinction curves also suggest that
dust enrichment by SNe II is indeed occurring at high-$z$
\citep{maiolino04,bianchi07,hirashita08}.
However, it is
not easy to explore the first dust enrichment in primeval
galaxies at high $z$ ($z\gtsim 5$) with present
observational facilities. Therefore, nearby templates of
primeval galaxies are useful to test galaxy formation
scenarios. The best
candidates for such a template are metal-poor
($\sim 1/10$--1/3 solar) blue compact dwarf
galaxies (BCDs),
since they are at the early stage of chemical evolution and
their typical stellar age is young
\citep{sargent70}. Indeed BCDs harbor appreciable
quantities of dust \citep{thuan99,plante02,hunt06}, and
can be used to investigate the dust properties in
chemically unevolved galaxies \citep{takeuchi_nozawa05}.

The dust content in a galaxy is often estimated by measuring
the dust emission in FIR. Because of the large all-sky
sample, the Infrared Astronomical Satellite (IRAS)
60~$\mu$m and 100 $\mu$m bands are most frequently used.
However, \citet{shibai99}, from the study of FIR spectral
energy distribution (SED) of the Milky Way, show that the IRAS
bands fail to detect a significant fraction of emission
from large grains which contribute
to wavelengths ($\lambda$) longer than 100 $\mu$m
(see also \cite{okumura02}). Moreover, with the size
distribution derived for the Milky Way dust grains,
the largest contribution to the dust content comes from
large grains ($a\sim 0.1~\mu$m, where $a$ is the grain
radius) \citep{mathis77}. Large grains also have
an importance that their temperature reflects the
stellar radiation field of interstellar medium through
the equilibrium between the heating from stellar
radiation and the radiative cooling in FIR
\citep{draine85}. In summary, FIR observation of
emission from large grains at
$\lambda >100~\mu$m is important to quantify the
dust content and the stellar radiation field.

Indeed, \citet{galliano05} show that some BCDs have a
cold dust component contributing to $\lambda >100~\mu$m.
\citet{popescu02} also point out that BCDs tend to have
the coldest dust temperature by using a Virgo cluster
sample observed by the Infrared Space Observatory (ISO),
although the sample size is small. This is unexpected
from the warm IRAS 60 $\mu$m/100 $\mu$m colors of BCDs
\citep{hoffman89}. \citet{engelbracht08}, by using the
70 $\mu$m and 160 $\mu$m bands of Multiband Imaging
Photometer for Spitzer
(MIPS) on the Spitzer Space Telescope, show that
the dust temperature
tends to rise as the metallicity decreases from solar
metallicity ($Z_\odot$) to $\sim 1/10~Z_\odot$. However,
there is a risk of contamination of stochastically heated
dust in the 70~$\mu$m band \citep{li01}, which leads to
overestimation of the large-grain temperature. Thus, a
statistical sample at $\lambda >100~\mu$m is indeed
required to
avoid possible contamination from stochastically heated
grains, and simultaneously to overcome the small number
statistics of
long-wavelength ($\lambda >100~\mu$m) observations.

In this paper, we report observations of BCDs in FIR
wavelengths including $\lambda > 100~\mu$m. The data are
obtained by AKARI \citep{murakami07} Far-Infrared
Surveyor (FIS) \citep{kawada07}. The continuous wavelength
coverage ($\lambda\sim 50$--180 $\mu$m) and 4 bands in
the FIR provided by FIS enable us not only to quantify the
long-wavelength ($\lambda >100~\mu$m) dust luminosity but
also to separate the contribution from stochastically
heated grains by using
the short-wavelength ($\lambda\ltsim 70~\mu$m) band. We
also demonstrate that this
advantage of FIS indeed makes it possible to
discuss some quantities related to dust properties and
dust enrichment.

This paper is organized as follows.
Observations and data reduction are described in
section \ref{sec:bcd}. Then, the basic data are
shown in section \ref{sec:result}, and some quantities
derived from those data are presented in section
\ref{sec:derived}. In section \ref{sec:discussion}
we discuss the results, and finally in
section \ref{sec:summary} we summarize this paper.

\section{Observations and Reduction}\label{sec:bcd}

The observations of eight BCDs are carried out by FIS
onboard AKARI as one of the open time observing
programs (Proposal ID: BCDDE, PI: H. Hirashita)
with four photometric bands of \textit{N60},
\textit{WIDE-S}, \textit{WIDE-L}, and \textit{N160},
whose central wavelengths are 65, 90, 140, and
160 $\mu$m with effective band widths of
$\Delta\lambda =21.7$, 37.9, 52.4, and 35.1 $\mu$m,
respectively \citep{kawada07}.
The measured FWHMs of the point spread function
(PSF) are
$37\pm 1''$, $39\pm 1''$, $58\pm 3''$, and
$61\pm 4''$, respectively, for the above four bands.

The sample BCDs are listed in Table \ref{tab:sample}.
We primarily selected the sample with IRAS detections as
listed in \citet{hirashita02} (originally compiled by
\cite{sage92} and \cite{lisenfeld98}). II~Zw~40, Mrk 7,
UM 439, UM 533, and II Zw 70 are finally observed.
The limitation to the sample comes from the lifetime and
visibility of the satellite. We also add Mrk 36 and
Mrk 71 from \citet{hopkins02} since they matched
the visibility requirement. Those galaxies are also
detected by IRAS. II~Zw~71 is also included in the sample,
since it lies in the field of view of II Zw 70. This is
the only galaxy which is not detected by IRAS. All the
sample BCDs are observed with the FIS01 scan sequence
(slow-scan observation for photometry),
whose total area scanned in the four bands is
$15'\times 12'$. Because of the faintness of the sample,
we choose the slower scan speed (8 arcsec/sec) and
the largest reset interval (2 sec).
The observational parameters are also
listed in Table \ref{tab:sample}.

\begin{table*}
  \caption{Observation log.}\label{tab:sample}
  \begin{center}
    \begin{tabular}{lcccccc}
      \hline
      Name     & R.A. (J2000.0) & Dec. (J2000.0) & Date & Observing mode &
               Scan speed & Reset interval
               \\ \hline
      II Zw 40 & \timeform{5h55m42s.6}  & \timeform{3D23'31''.8}  &
                2006 Sep.\ 21 & FIS01 & 8$''~\textrm{s}^{-1}$ & 2 s\\
      Mrk 7    & \timeform{7h28m11s.4}  & \timeform{72D34'23''.4} &
                2006 Oct.\ 3  & FIS01 & 8$''~\textrm{s}^{-1}$ & 2 s\\
      Mrk 71   & \timeform{7h28m42s.8} & \timeform{69D11m21''.4}  &
                2006 Oct.\ 4  & FIS01 & 8$''~\textrm{s}^{-1}$ & 2 s\\
      UM 439   & \timeform{11h36m36s.8} & \timeform{0D48'58''.0}  &
                2006 Dec.\ 16 & FIS01 & 8$''~\textrm{s}^{-1}$ & 2 s\\
      UM 533   & \timeform{12h59m58s.1} & \timeform{2D02'57''.3}  &
                2007 Jan.\ 4  & FIS01 & 8$''~\textrm{s}^{-1}$ & 2 s\\
      II Zw 70\footnotemark[$*$] &
                 \timeform{14h50m56s.5} & \timeform{35D34'17''.6} &
                2007 Jan.\ 15 & FIS01 & 8$''~\textrm{s}^{-1}$ & 2 s\\
      II Zw 71\footnotemark[$*$] &
                 \timeform{14h51m14s.4} & \timeform{35D32'32''.2} &
                2007 Jan.\ 15 & FIS01 & 8$''~\textrm{s}^{-1}$ & 2 s\\
      Mrk 36   & \timeform{11h04m58s.5} & \timeform{29D08'22''.1} &
                2007 May 27   & FIS01 & 8$''~\textrm{s}^{-1}$ & 2 s\\
      \hline
       \multicolumn{4}{@{}l@{}}{\hbox to 0pt{\parbox{85mm}{\footnotesize
       \par\noindent
       \footnotemark[$*$] II Zw 70 and II Zw 71 lie in the same field of view.
    }\hss}}    
    \end{tabular}
  \end{center}
\end{table*}

The raw data were reduced by using the FIS Slow Scan Tool
(version
20070914).\footnote{http://www.ir.isas.ac.jp/ASTRO-F/Observation/.}
The absolute calibration of the FIS slow-scan data is
based on the
COBE/DIRBE\footnote{http://cmbdata.gsfc.nasa.gov/product/cobe/.}
measurements of the diffuse sky emission from zodiacal
light and interstellar cirrus averaged over areas of
several slow-scan observations. Because the detector response
is largely affected by a hit of high energy ionizing particle,
we use the local flat; that is, we correct the detector
sensitivities by assuming uniformity of the sky brightness
\citep{verdugo07}.\footnote{We referred to AKARI FIS Data User
Manual Version 2 available at
http://www.ir.isas.ac.jp/ASTRO-F/Observation/. When we run the
pipeline command {\tt ss\_run\_ss}, we apply options
{\tt /local,/smooth,width\_filter=80}:
The flat field is built from the observed sky,
and boxcar smoothing with a filter width of 80 s in
the time series data is applied to remove
remaining background offsets among the pixels.}
By using some relatively bright sources, we confirmed that the
errors caused by the smoothing procedures are well within
10\% for \textit{N60} ($65~\mu$m) and
\textit{WIDE-S} ($90~\mu$m) and 20\% for
\textit{WIDE-L} ($140~\mu$m). These error levels are
consistent to or
somewhat larger than the background fluctuation
mainly caused by a hit of high energy ionizing particle,
which implies that the background fluctuation is a major
component in the errors. The above values are adopted for
the errors in
Table \ref{tab:sample} except for the following two cases:
(i) For the \textit{N60} ($65~\mu$m) flux of Mrk 7,
we adopt the root-mean-square (rms) of the
background fluctuation. (ii) The uncertainty in the
\textit{WIDE-S} ($90~\mu$m) data of II Zw 70 is also large
because of a bad pixel within the aperture radius
(section \ref{subsec:remark}). For \textit{N160} ($160~\mu$m),
because of
high fluctuation of the background, we adopt the 25\% error
after estimating the uncertainty in the background subtraction.
When the flux is smaller than the rms of the
background,
we adopt 3 times the background uncertainty as an upper limit.

Our sample BCDs are compact enough to be treated as point
sources with the resolution of FIS (but see
section \ref{subsec:remark} for Mrk 71). We follow the
aperture photometry described in \citet{verdugo07}. We adopt
aperture radii of 0.625 arcmin for \textit{N60} ($60~\mu$m)
and \textit{WIDE-S} ($90~\mu$m) and of 0.750 arcmin for
\textit{WIDE-L} ($140~\mu$m) and \textit{N160} ($160~\mu$m),
and sky regions between 2.25 and 3.25 arcmin for \textit{N60}
($65~\mu$m) and \textit{WIDE-S} ($90~\mu$m) and between
3.0 and 4.0 arcmin for \textit{WIDE-L} ($140~\mu$m)
and \textit{N160} ($160~\mu$m). Then the aperture correction
factors (1.58, 1.74, 1.71, and 2.03
for \textit{N60}, \textit{WIDE-S}, \textit{WIDE-L},
and \textit{N160}, respectively)
are multiplied to obtain the total flux.\footnote{We also multiplied
another factor called correction factor. It is an empirical
factor to correct the discrepancy between the calibration of
the diffuse emission (on which the present reduction is based) and
that for point sources.
This correction factor is empirically known to be constant.
By using a different tool developed by T. Suzuki, who adopt
point source calibrations by asteroids, we confirmed that
the correction factors are reasonable within the errors put
in this paper. See also \citet{kawada07} for the absolute
calibration.}

Finally, color corrections are applied for the
\textit{WIDE-L} ($140~\mu$m) fluxes: the color correction
factor is assumed to be 0.93 (the flux is divided by this
factors), which is valid
for the temperature range derived by using 140 $\mu$m
and 90 $\mu$m fluxes (22--53 K for $\beta =1$ and
19--36 K for $\beta =2$; see section \ref{subsec:Tdust}) within
3\%.  For the \textit{WIDE-S} ($90~\mu$m)
band, color correction would be required. The
correction factor is 0.94 at 35 K and 1.06 at 50 K
for $\beta =1$, and 0.94 at 30 K and 1.06 at 40 K.
These temperature ranges cover the average of
140 $\mu$m--90 $\mu$m temperature and
65 $\mu$m--90 $\mu$m temperature
(section \ref{subsec:Tdust}). For II Zw 71, whose
temperature is significantly lower than the above ranges,
upward $\sim 10\%$ correction would be required.
However, we do not apply color correction to the
\textit{WIDE-S} ($90~\mu$m) flux, because the
emission in this band may not be described with a
single-temperature emission and because the correction
factor is not uniform within the sample (i.e., an
iteration between the derived dust temperature and
color correction is required for each object).
For \textit{N60} (65 $\mu$m) and \textit{N160} (160 $\mu$m),
we do not apply
color correction, since the color correction change
the flux only $<4$\% for our sample within the
range of dust temperature derived for our sample in
section \ref{subsec:Tdust}.

\section{Results}\label{sec:result}

\subsection{Flux}

In Table \ref{tab:flux}, we present the measured fluxes.
All the BCDs are detected in \textit{N60} ($65~\mu$m),
\textit{WIDE-S} ($90~\mu$m), and \textit{WIDE-L}
($140~\mu$m) bands except for Mrk 36
(section \ref{subsec:remark}). For the \textit{N160}
($160~\mu$m) band, a half of the
objects are detected and only upper limits are
determined for the other half.
For comparison, in Table \ref{tab:flux} we also compile
the IRAS 60~$\mu$m and
100 $\mu$m fluxes if available. We adopt those data from
\citet{moshir90} except for II Zw 40,
Mrk 71 and II Zw 71 (section \ref{subsec:remark}). For II Zw 40,
we also present the
data taken by Spitzer MIPS at $\lambda =70~\mu$m
and 160 $\mu$m \citep{engelbracht08}.

\begin{table*}
  \caption{Measured fluxes.}\label{tab:flux}
  \begin{center}
    \begin{tabular}{lcccc|cc}
      \hline
      Name     & \textit{N60} & \textit{WIDE-S}  &
               \textit{WIDE-L} & \textit{N160} &
               IRAS & IRAS
               \\
               & ($\lambda =65~\mu$m) & ($\lambda =90~\mu$m) &
               ($\lambda =140~\mu$m) & ($\lambda =160~\mu$m) &
               ($\lambda =60~\mu$m) & ($\lambda =100~\mu$m) \\ \hline 
      II\,Zw\,40\footnotemark[$*$] & $6.9\pm 0.7$\,Jy & $6.6\pm 0.7$\,Jy &
               $3.7\pm 0.7$\,Jy & $3.4\pm 0.9$\,Jy & $6.61\pm 0.7$\,Jy
               & $5.8\pm 0.9$\,Jy \\
      Mrk 7    & $0.83\pm 0.23$\,Jy & $0.84\pm 0.08$\,Jy &
               $1.1\pm 0.2$\,Jy & $1.0\pm 0.3$\,Jy & $0.48\pm 0.04$\,Jy
               & $0.97\pm 0.14$\,Jy \\
      Mrk 71\footnotemark[$\dagger$]   & $3.7\pm 0.4$\,Jy & $3.1\pm 0.3$\,Jy &
               $2.2\pm 0.4$\,Jy & $<3.0$\,Jy & $4.83\pm 0.04$\,Jy &
               $5.71\pm 0.03$\,Jy \\
      UM\,439   & $0.42\pm 0.04$\,Jy & $0.40\pm 0.04$\,Jy &
               $0.31\pm 0.06$\,Jy & $<0.45$\,Jy & $0.36\pm 0.05$\,Jy
               & $0.81\pm 0.11$\,Jy \\
      UM\,533   & $0.54\pm 0.05$\,Jy & $0.61\pm 0.06$\,Jy &
               $0.54\pm 0.11$\,Jy & $<0.68$\,Jy & $0.50\pm 0.05$\,Jy
               & $0.54\pm 0.13$\,Jy \\
      II\,Zw\,70 & $0.89\pm 0.09$\,Jy & $0.80\pm 0.24$\,Jy &
               $0.75\pm 0.15$\,Jy & $0.62\pm 0.16$\,Jy &
               $0.71\pm 0.05$\,Jy & $1.24\pm 0.12$\,Jy \\
      II\,Zw\,71 & $0.32\pm 0.03$\,Jy & $0.54\pm 0.05$\,Jy&
               $1.3\pm 0.26$\,Jy & $1.3\pm 0.3$\,Jy & --- & --- \\
      Mrk 36   & $0.33\pm 0.03$\,Jy & $0.26\pm 0.03$\,Jy &
               $<0.30$ Jy & $<0.38$\,Jy & $0.23\pm 0.06$\,Jy &
               $0.68\pm 0.08$\,Jy \\
      \hline
       \multicolumn{7}{@{}l@{}}{\hbox to 0pt{\parbox{170mm}{\footnotesize
       Note. 3 $\sigma$ upper limits are shown by ``$<$''
       (For the \textit{WIDE-L} data of Mrk 36, see
       section \ref{subsec:remark}). ``---''
       indicates no data. IRAS 60 $\mu$m and 100 $\mu$m data are also
       shown for comparison in the last two columns.
       \par\noindent
       \footnotemark[$*$] II Zw 40 has also Spitzer MIPS data
       ($5.58\pm 0.28$ Jy at $\lambda =70~\mu$m and $3.14\pm 0.43$ Jy
       at $\lambda =160~\mu$m; \cite{engelbracht08}).
       \par\noindent
       \footnotemark[$\dagger$] The discrepancy
       between our results and the IRAS data may be explained by different
       resolution (section \ref{subsec:remark}).
    }\hss}}
    \end{tabular}
  \end{center}
\end{table*}

\subsection{FIR Spectral Energy Distributions}
\label{subsec:sed}

\begin{figure*}
  \begin{center}
    \FigureFile(60mm,60mm){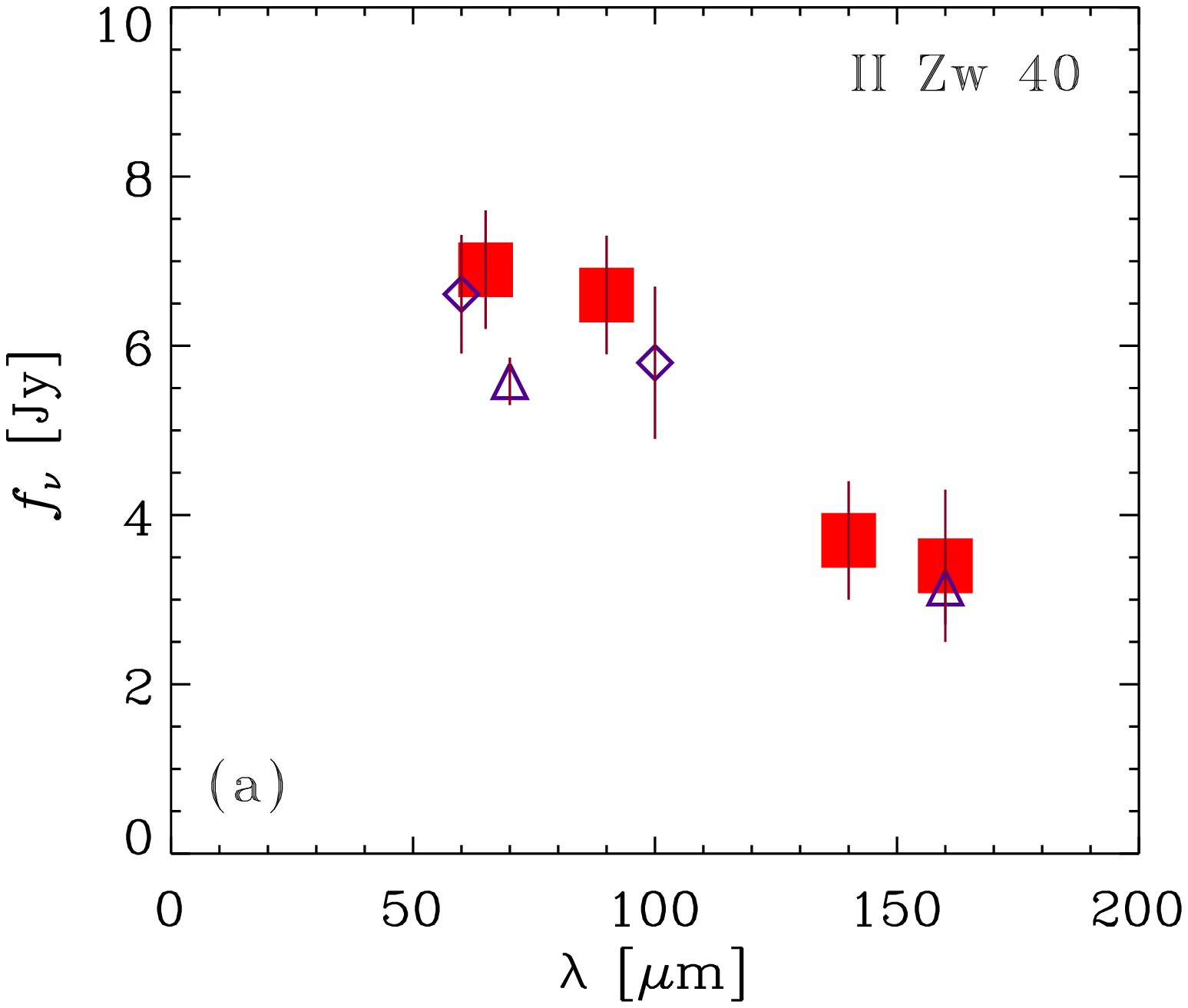}
    \FigureFile(60mm,60mm){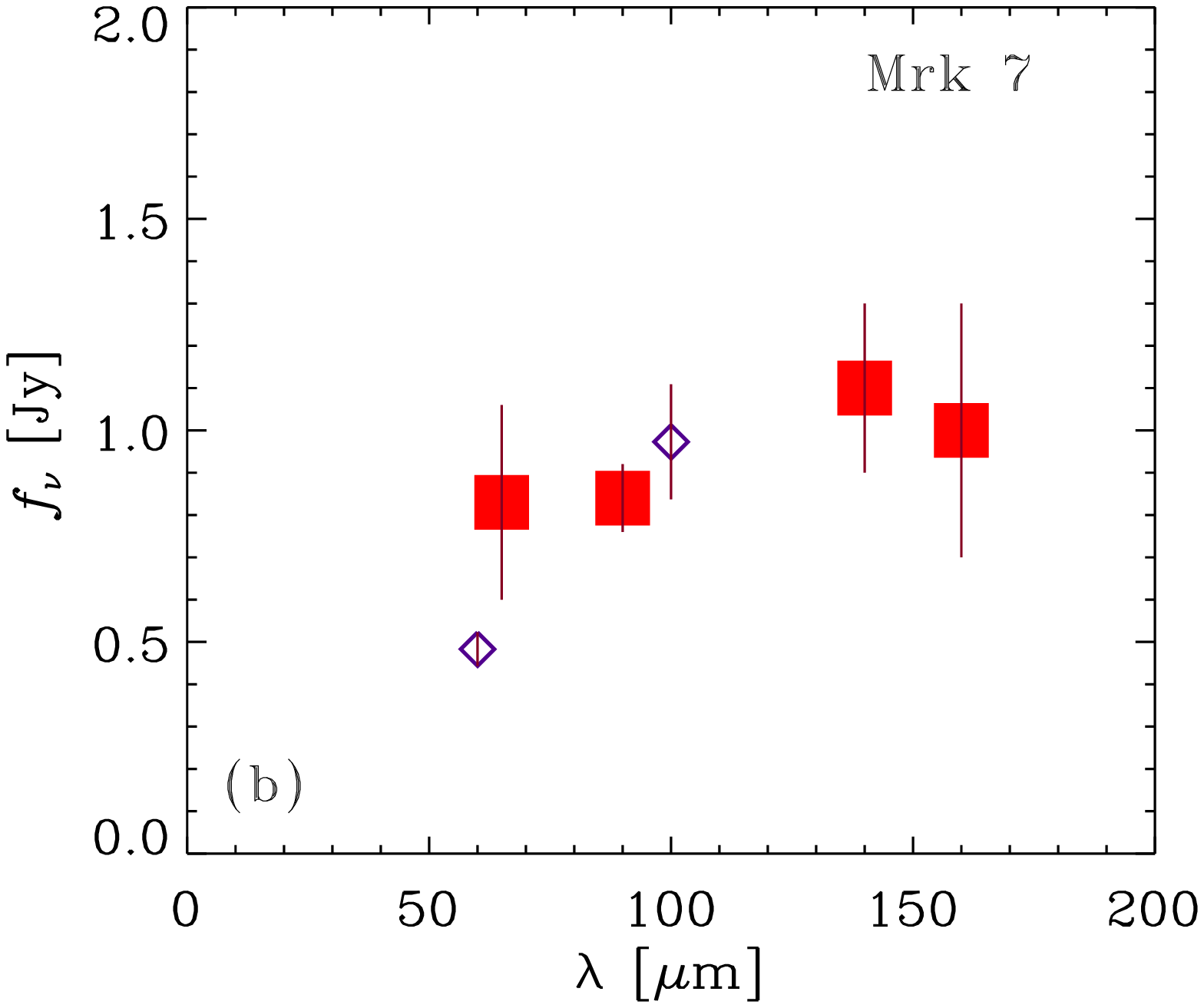}
    \FigureFile(60mm,60mm){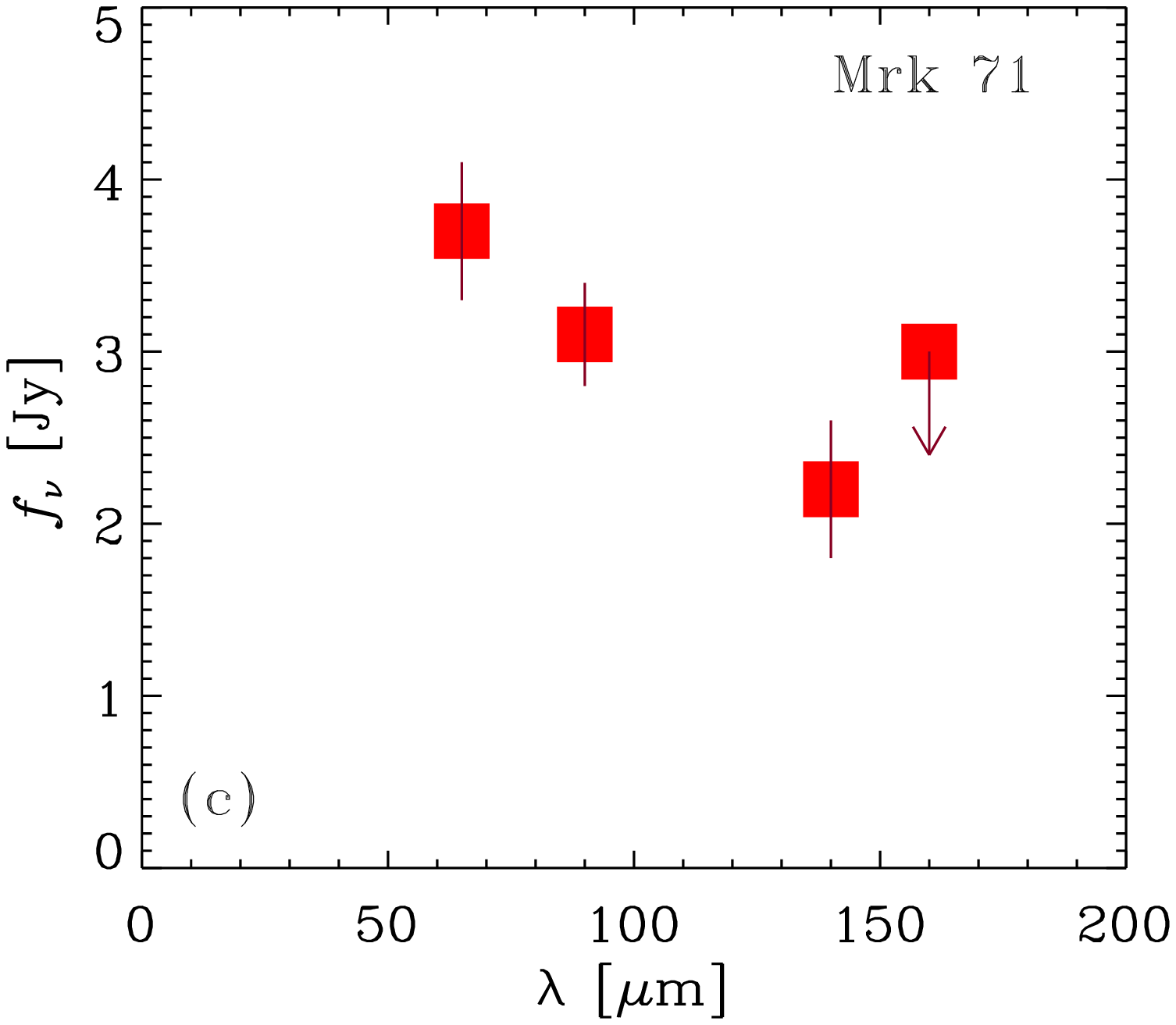}
    \FigureFile(60mm,60mm){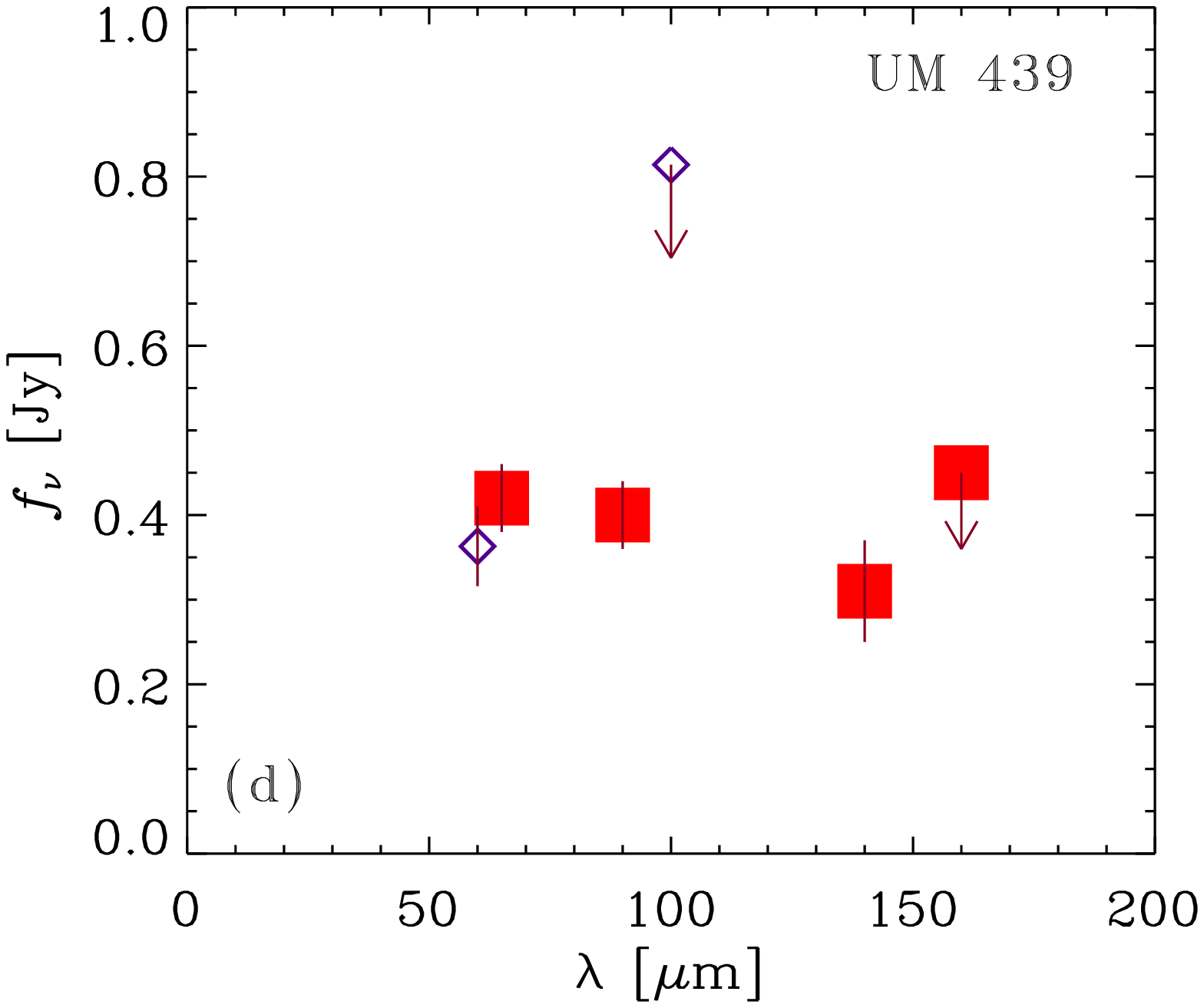}
    \FigureFile(60mm,60mm){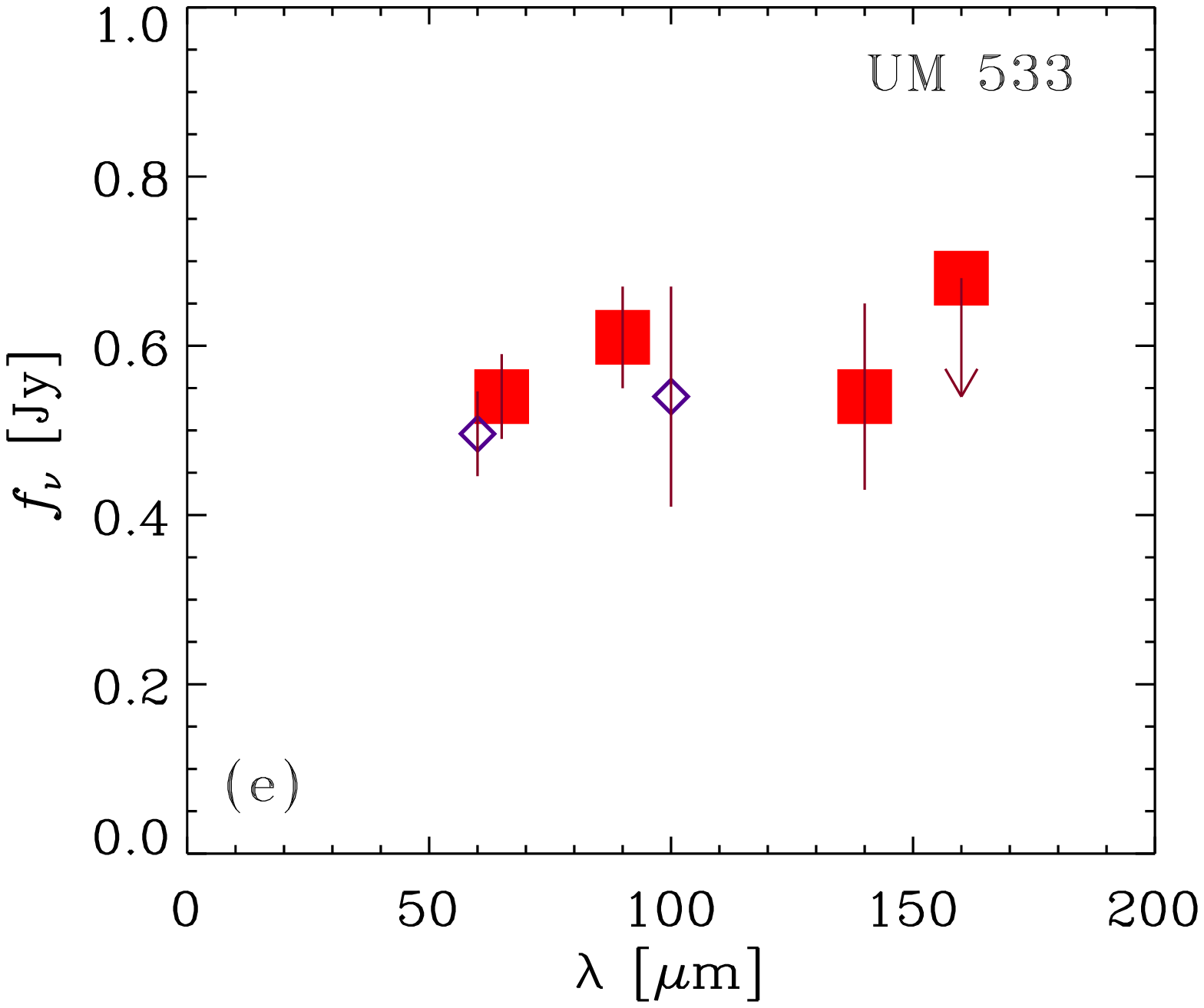}
    \FigureFile(60mm,60mm){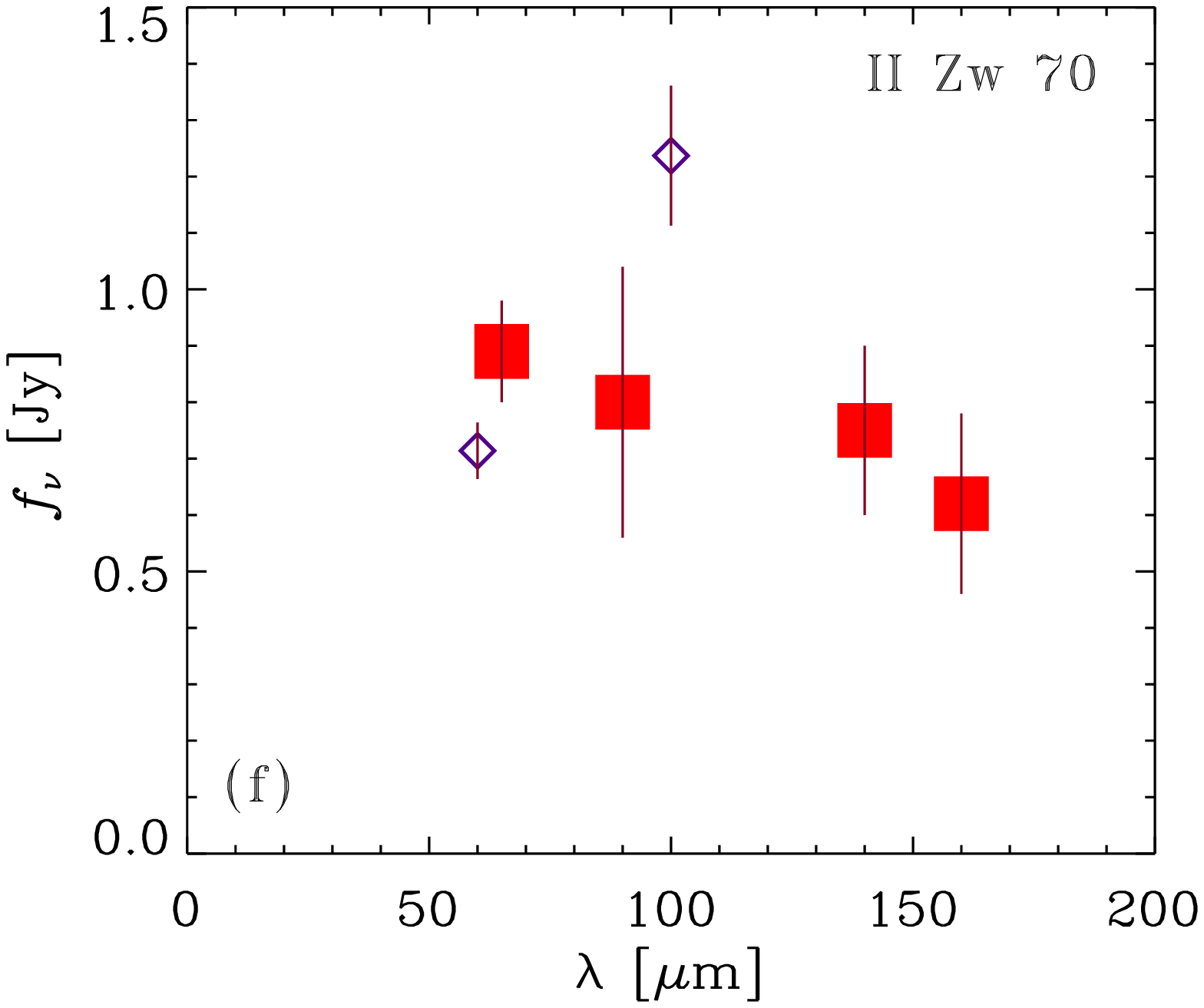}
    \FigureFile(60mm,60mm){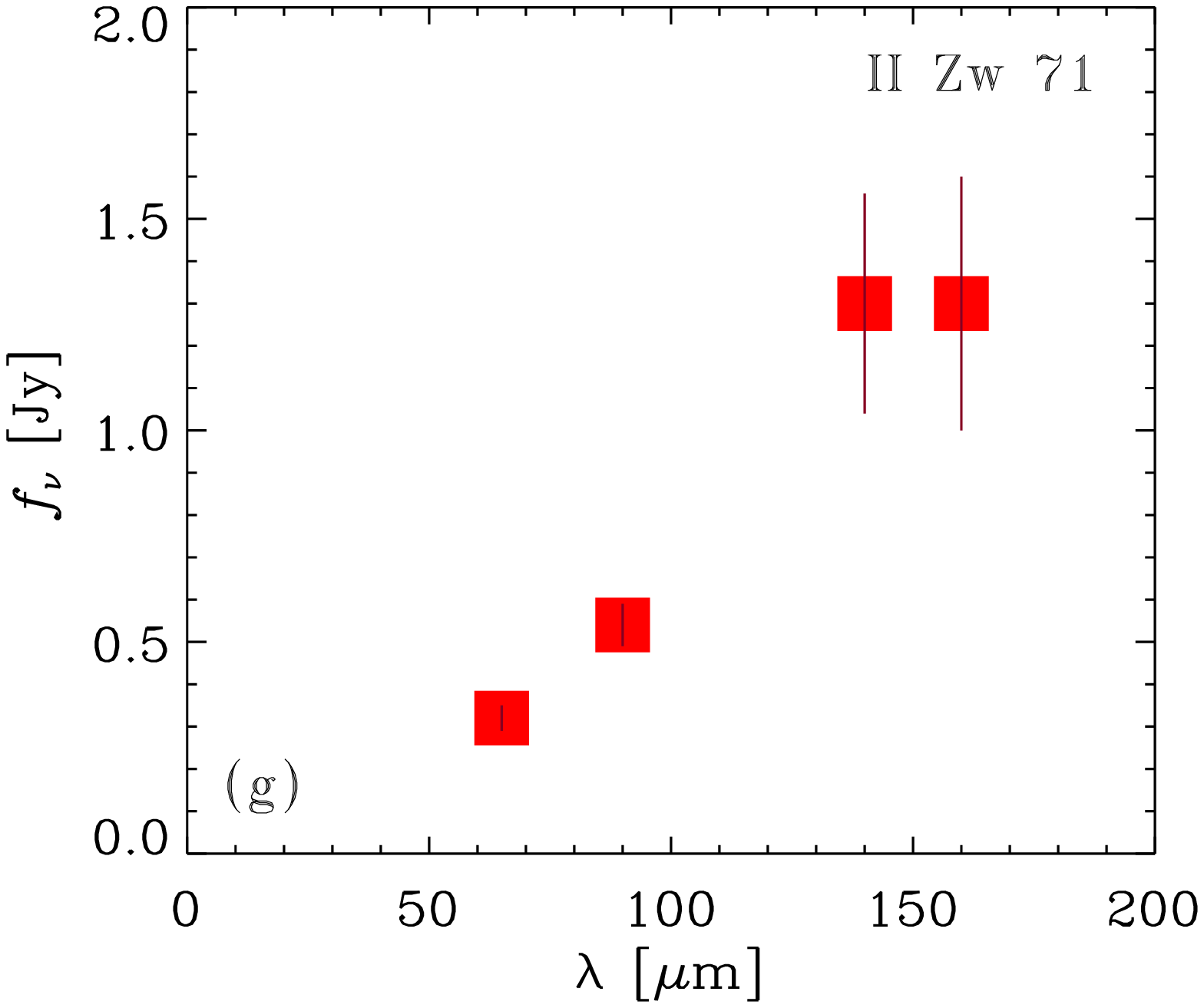}
    \FigureFile(60mm,60mm){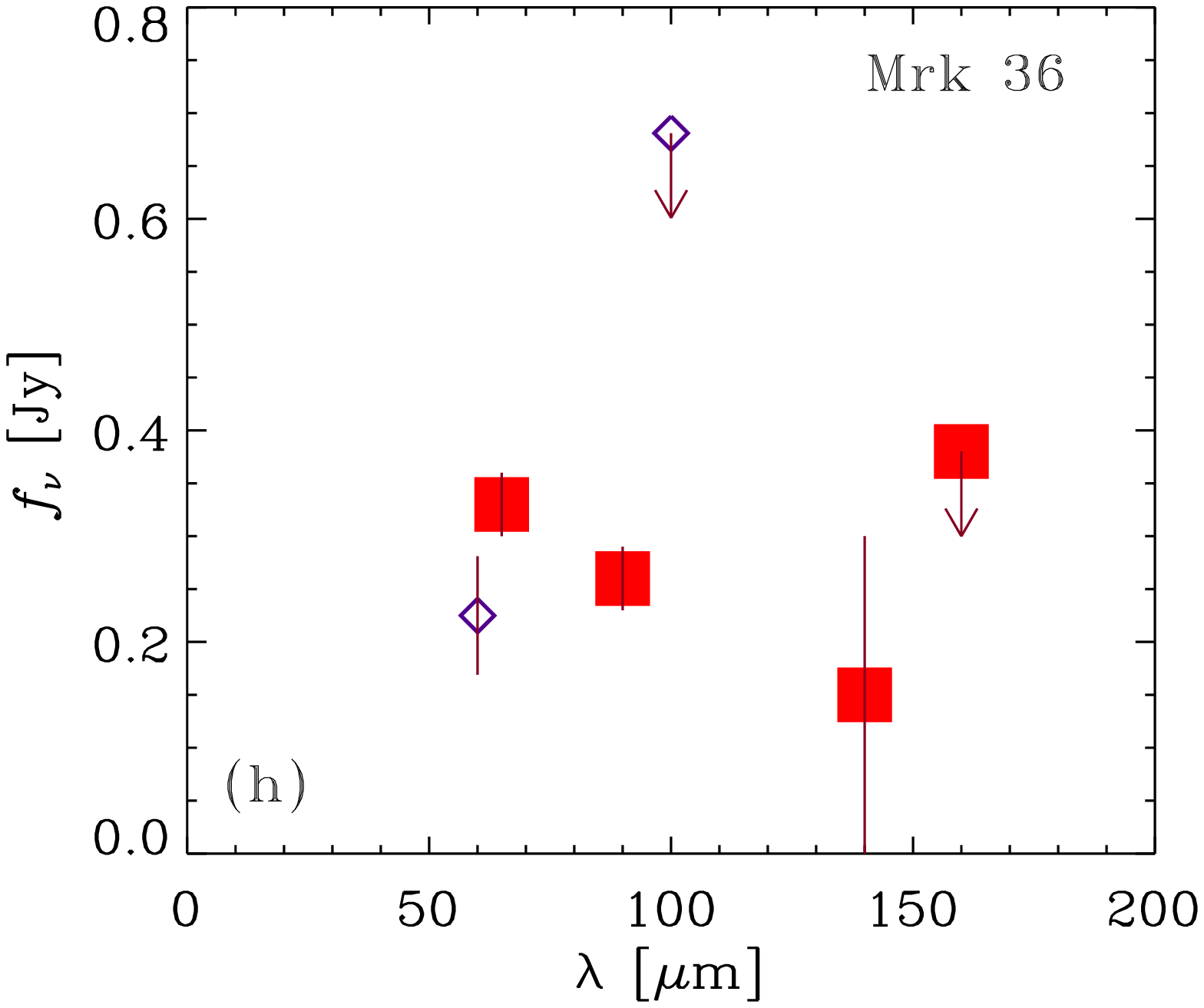}
  \end{center}
  \caption{Observed FIR SEDs at various bands
  ($f_\nu$ is flux in units of Jy and $\lambda$ is
  wavelength in units of $\mu$m).
  AKARI data are shown by filled squares and IRAS data
  are also presented by open diamonds
  for comparison. For II Zw 40, Spitzer data are
available and are shown by open triangles.
The bars indicate errors and the points
  with arrows give 3 $\sigma$ upper limits. The name
  of object is shown in each panel.}
  \label{fig:sample}
\end{figure*}

The data in Table \ref{tab:flux} are displayed in
Figure \ref{fig:sample}. For almost all the sources with
available IRAS data, the AKARI and IRAS fluxes are
consistent but there is a significant discrepancy between
the IRAS 100 $\mu$m flux ($1.24\pm 0.12$~Jy) and the
AKARI 90 $\mu$m ($0.80\pm 0.24$~Jy) flux for II Zw 70.
Note that the error bar for the 90~$\mu$m flux is large
because of the bad pixel
(section \ref{sec:bcd}). However, we should note that
the discrepancy seems significant even after considering
the uncertainty in the interpolation.
Another possible reason is that the IRAS flux
is sometimes overestimated \citep{jeong07}. Source
confusion could explain the discrepancy, but there is
no evidence for a large sky fluctuation around II Zw 70
in our image.

The \textit{N160} ($160~\mu$m) flux, when detected, is
always near to the \textit{WIDE-L} ($140~\mu$m).
Thus, the 160 $\mu$m flux does not provide strongly
independent information of the 140 $\mu$m flux.
Because of this and of a large fraction of non-detection,
we do not use 160 $\mu$m data in the following discussions.
Nevertheless, it is worth stressing that the consistency
between the 140 $\mu$m and 160 $\mu$m fluxes confirms
the reliability of the \textit{N160} band.
The \textit{N160} flux of II Zw 40 is also consistent with
the Spitzer 160 $\mu$m flux, which also strengthen the
reliability of the \textit{N160} fluxes.

\subsection{Remarks on individual objects}
\label{subsec:remark}

\noindent
II Zw 40 --- The IRAS data are taken from \citet{vader93}
(see also \cite{beck02}). The Spitzer MIPS data are also
available in \citet{engelbracht08}.

\citet{galliano05} investigate this galaxy by using detailed
SED models. They suggest a very cold dust contributing to the
emission at $\lambda\gtsim 300~\mu$m (see also
sections \ref{subsec:Mdust} and \ref{subsec:dust_metal}).
However, we cannot constrain the dust emission at such a long
wavelength range. \citet{hunt05} also study the SED over a
wide wavelength range including FIR, and state that the FIR
SED has a peak at $\lambda\ltsim 60~\mu$m, which is
consistent with our data. This galaxy has the strongest
interstellar radiation field, which supports high density of
star formation \citep{beck02,vanzi08}.

\vspace{3mm}



\noindent
Mrk 71 ---
This is called ``cometary'' BCD \citep{noeske00} and given
an NGC number NGC 2366. This is the nearest BCD of our
sample with distance of 3.4 Mpc
\citep{tolstoy95}. Indeed, an extended structure in
the north-east direction from the main H {\sc ii}
complex NGC 2363 is seen in optical images. Since
this paper
treats the main star-forming region of a BCD,
we concentrate on the main complex, which can be
treated as a point source with the resolution of
FIS. The north-east
extension is detected in the \textit{N60}
($65~\mu$m),
\textit{WIDE-S} ($90~\mu$m), and
\textit{WIDE-L} ($140~\mu$m) band images.
This extended structure does not
affect our analysis because of its low surface
brightness within the aperture used for the
photometry.

With IRAS data, \citet{mazzarella91} derived larger
FIR fluxes (4.829 Jy at $\lambda =60~\mu$m and
5.713 Jy at $\lambda =100~\mu$m) than our estimate.
In the north-west extension, there is a brightness
peak at distance of $\sim 3'$ from the main body.
The flux around this peak is preliminarily estimated
as 0.7 Jy at $\lambda =90~\mu$m and
0.5 Jy at $\lambda =65~\mu$m.
This peak can be contaminated with the main body
with the resolution of IRAS ($\sim 2'$--5$'$)
\citep{neugebauer84}.
Thus, we do not
compare AKARI and IRAS results for this galaxy
(section \ref{subsec:sed}). A
quantitative and detailed analysis of the north-east
extension will be reported in a separate paper.

\vspace{3mm}




\vspace{3mm}

\noindent
II Zw 70 ---
There is a bad pixel within the photometric radius in
\textit{WIDE-S}. Although the brightness of the pixel is
extrapolated by using the value of the surrounding
pixels, the uncertainty caused by this extrapolation
may be as large as 30\%.\footnote{We derived
this error by using another analysis tool developed
by \citet{suzuki08}.} Thus, we put a 30\% error bar
to the \textit{WIDE-S} (90 $\mu$m) flux.

\vspace{3mm}

\noindent
II Zw 71 ---
This galaxy is occasionally detected in the same
field of II Zw 70. It is located at the edge of the
field of view, but the aperture used
to determine the flux is well within the field.
Although the area for
the background is smaller than the other sources,
the number of background pixels is
large enough to subtract the background
within errors of 10\% for \textit{N60}
($65~\mu$m) and \textit{WIDE-S}
($90~\mu$m), 20\% for \textit{WIDE-L}
($140~\mu$m), and 25\% for
\textit{N160} ($160~\mu$m).
This galaxy is not detected by IRAS.

\vspace{3mm}

\noindent
Mrk 36 ---
This is the faintest object in \textit{WIDE-S}.
This galaxy is detected by IRAS at $\lambda =60~\mu$m
with a flux level of $0.022\pm 0.05$ Jy and is not
detected at
$\lambda =100~\mu$m with an upper limit of
$0.68$ Jy \citep{moshir90}.
In the \textit{WIDE-L} image, it seems
that relatively bright pixels are clustered in the expected
position of Mrk 36.
If we use those clustered bright pixels
and subtract the background, the flux is 0.15 Jy.
But since the uncertainty in the background is
comparable to this value, it is safe to put an upper limit
of 0.30 Jy.

\section{Derived Quantities}\label{sec:derived}

\subsection{Dust temperature}\label{subsec:Tdust}

One of the most straightforward quantities derived by
our observations is colors (i.e., flux ratios between two
bands), which reflect dust temperatures.
In Table \ref{tab:Tdust}, we show the colors.
Since the \textit{WIDE-S} (90 $\mu$m) band is the most
sensitive, we present the color relative to
90~$\mu$m flux. We denote the flux ratio between
wavelengths $\lambda_1$~[$\mu$m] and
$\lambda_2$ [$\mu$m] as [$\lambda_1$, $\lambda_2$].
The error of a flux ratio $f_1/f_2$, where we denote the
errors of $f_1$ and $f_2$ as err($f_1$) and err($f_2$),
respectively, is estimated
by using a simple discussion of error propagation as
\begin{eqnarray}
\mathrm{ err}\left(\frac{f_1}{f_2}\right) =
\frac{f_1}{f_2}\sqrt{\frac{\mathrm{ err}(f_1)^2}{f_1^2}+
\frac{\mathrm{ err}(f_2)^2}{f_2^2}\,}\, .
\end{eqnarray}

\begin{table*}
  \caption{Flux ratios and dust temperatures.}\label{tab:Tdust}
  \begin{center}
    \begin{tabular}{lcccccccc}
      \hline
      Name     & [65/90] & [140/90]  & $T_\mathrm{ d}([65/90];\, 1)$
               & $T_\mathrm{ d}([140/90]; 1)$ & $T_\mathrm{ d}([65/90];\, 2)$
               & $T_\mathrm{ d}([140/90];\, 2)$
                              \\ \hline
      II Zw 40 & $1.0\pm 0.15$ & $0.56\pm 0.12$ &
               $50^{+7}_{-6}$ K & $53^{+20}_{-9}$ K & $39^{+4}_{-3}$ K &
               $36^{+8}_{-4}$ K \\
      Mrk 7    & $0.99\pm 0.29$ & $1.3\pm 0.3$ &
               $48^{+13}_{-11}$ K & $28^{+4}_{-2}$ K & $38^{+7}_{-7}$ K
               & $23^{+3}_{-1}$ K \\
      Mrk 71   & $1.2\pm 0.2$ & $0.71\pm 0.15$ & $57^{+9}_{-8}$ K
               & $42^{+10}_{-5}$ K & $43^{+5}_{-4}$ K & $31^{+5}_{-3}$ K
               \\
      UM 439   & $1.1\pm 0.15$ & $0.78\pm 0.17$ &
               $50^{+7}_{-5}$ K & $39^{+10}_{-5}$ K & $39^{+4}_{-3}$ K
               & $30^{+4}_{-3}$ K \\
      UM 533   & $0.89\pm 0.12$ & $0.89\pm 0.20$ &
               $44^{+5}_{-4}$ K & $36^{+7}_{-5}$ K & $35^{+3}_{-2}$ K &
               $28^{+4}_{-3}$ K \\
      II Zw 70 & $1.1\pm 0.35$ & $0.94\pm 0.34$ &
               $53^{+19}_{-14}$ K & $34^{+15}_{-5}$ K & $41^{+10}_{-9}$ K
               & $27^{+8}_{-4}$ K \\
      II Zw 71 & $0.59\pm 0.08$ & $2.4\pm 0.5$ &
               $34^{+2}_{-2}$ K & $22^{+2}_{-2}$ K & $29^{+1}_{-2}$ K &
               $19^{+1}_{-2}$ K \\
      Mrk 36   & $1.3\pm 0.2$ & $0.58\pm 0.58$ &
               $61^{+11}_{-9}$ K & $>30$ K & $45^{+6}_{-5}$ K &
               $>24$ K \\ \hline
      NGC 2841 & $0.29\pm 0.08$ & $1.9\pm 0.7$ &
               $24^{+2}_{-3}$ K & $24^{+6}_{-3}$ K & $21^{+2}_{-2}$ K &
               $20^{+4}_{-2}$ K \\
      NGC 2976 & $0.43\pm 0.12$ & $1.14\pm 0.41$ &
               $29^{+3}_{-4}$ K & $31^{+10}_{-5}$ K & $25^{+3}_{-3}$ K &
               $25^{+6}_{-3}$ K \\
      M 101    & $0.36\pm 0.10$ & $2.1\pm 0.8$ &
               $27^{+3}_{-4}$ K & $23^{+5}_{-3}$ K & $23^{+3}_{-2}$ K &
               $19^{+4}_{-1}$ K \\
      \hline
       \multicolumn{7}{@{}l@{}}{\hbox to 0pt{\parbox{180mm}{\footnotesize
       Note. Lower limits are shown by ``$>$''. The last three galaxies
             are spiral galaxies for comparison.
             The data are taken from \citet{kaneda07} for NGC 2841 and
             NGC~2976 and from \citet{suzuki07} for M 101.
    }\hss}}
    \end{tabular}
  \end{center}
\end{table*}

We compare the derived colors with other available
samples of star-forming galaxies observed by AKARI FIS.
To date,
FIS observations of three spiral galaxies have been
published to date \citep{kaneda07,suzuki07}
as shown in Table \ref{tab:Tdust}. We observe
that the derived colors of the
BCDs show higher temperatures
(i.e., larger [65,~90] and smaller [140,~90])
than the spiral galaxies
except for Mrk 7 and II Zw 71. This supports the idea that
BCDs host intense star formation activities in
concentrated regions.

We have selected the present sample primarily from the
IRAS catalog. Since intense star formation tends to
raise FIR luminosity, our sample may be biased to
objects with intense star formation. Indeed,
II Zw 71, which is not detected by IRAS,
has the lowest dust temperature, and this fact implies
a possibility that
our selection by IRAS detection is biased to high dust
temperature.

Now we estimate the dust temperature from the flux ratio.
We fit $A\nu^{\beta}B_\nu (T_\mathrm{d})$ ($A$ is a
constant, $B_\nu (T)$ is the Planck function evaluated
at frequency $\nu$ and temperature $T$ and $\beta$ is
the index of the emission coefficient called emissivity
index) to the fluxes at two bands to derive the dust
temperature $T_\mathrm{d}$. Note that $A$ can be eliminated
if we take a flux ratio.
In Table \ref{tab:Tdust}, we show the estimated
dust temperature, where $T_\mathrm{d}(C;\,\beta)$
represents the dust temperature derived from color
$C$ and emissivity index $\beta$.

{}From Table \ref{tab:Tdust}, we observe that the values
of $T_\mathrm{d}$ is higher for $\beta =1$ than for
$\beta =2$, because under fixed fluxes at two wavelengths
$\beta =2$ requires lower $T_\mathrm{d}$ (i.e., enhanced
$B_\nu (T_\mathrm{d})$ at the longer wavelength) to
compensate for lower emission coefficient at the longer
wavelength. Again we confirm that the dust temperature
of the BCDs is significantly higher than that of the
spiral sample except for Mrk~7 and II Zw 71. We also see
that there is a large variety of dust temperatures among
our sample. Such a large variety of dust temperatures
is also reported by \citet{engelbracht08} for a
low-metallicity sample. One of their sample,
II Zw 40 is common with \citet{engelbracht08}, and the
dust temperature derived by them by using the
Spitzer 70 and 160 $\mu$m bands with
$\beta =2$ is $33.4\pm 1.0$ K, consistent with our
$T_\mathrm{ d}([140/90]; \, \beta =2)=36^{+8}_{-4}$ K.

If $T_\mathrm{ d}([65/90])$ is equal to
$T_\mathrm{ d}([140/90])$,\footnote{When the parameter
$\beta$ is omitted, the statement is true for any
$\beta$.}
the fluxes in the three bands can be fitted with
a single dust temperature. The difference between
$T_\mathrm{ d}([65/90])$ and $T_\mathrm{ d}([140/90])$
is significant for Mrk 7 and II Zw 71 if $\beta =1$, and
for Mrk 7, Mrk 71, UM 439, UM 533, and II Zw 71 if
$\beta =2$. In those BCDs, $T_\mathrm{d}([65/90])$ is
higher than
$T_\mathrm{ d}([140/90])$, which possibly indicates that
the contribution from
stochastically heated grains to the 65 $\mu$m flux is
prominent compared with the short-wavelength tail of
the emission at $\lambda >90~\mu$m \citep{draine01}.
Since the emission at $\lambda >90~\mu$m comes from
dust grains in temperature equilibrium between the
heating from stellar radiation field (called
interstellar radiation field)
and the cooling by FIR reemission,
observations by the \textit{WIDE-S} and \textit{WIDE-L}
bands are essential to estimate the equilibrium grain
temperature. In such a case,
the dust temperature derived from the IRAS 60~$\mu$m
and 100~$\mu$m bands is overestimate for the
equilibrium grain temperature. For BCDs in which
$T_\mathrm{ d}([65/90])$ is consistent with
$T_\mathrm{ d}([140/90])$, on the contrary, the IRAS
60 $\mu$m and 100 $\mu$m bands have given
reasonable estimate for the equilibrium
dust temperature. In this case, the Spitzer
70 $\mu$m and 160 $\mu$m bands also provide
the equilibrium grain temperature.

It is worth noting that as demonstrated above the AKARI
FIS bands have the following two advantages: One is
that two bands, \textit{WIDE-S} ($90~\mu$m) and
\textit{WIDE-L} ($140~\mu$m), are available to
estimate the
equilibrium grain temperature, while the Spitzer
$70~\mu$m and 160 $\mu$m bands have risk of
overestimating the equilibrium grain temperature
because of the contribution from stochastically heated
grains to $\lambda =70~\mu$m.
(But a wide band has a disadvantage in
determining the flux density. In fact the color
correction factor is relatively large for
\textit{WIDE-S} (90~$\mu$m) and
\textit{WIDE-L} (140~$\mu$m) as
stated in section \ref{sec:bcd}.)
The other is that the
\textit{N60} band ($65~\mu$m) is also crucial
to estimate the contribution from stochastically heated
grains. Thus, the three bands of FIS adopted in this
paper provide a unique opportunity to investigate
\textit{both} equilibrium-temperature dust and
stochastically heated dust.

\subsection{Interstellar radiation field}
\label{subsec:ISRF}

Since the dust temperature is determined by the
equilibrium between the heating from interstellar
radiation field and the cooling by FIR reemission, it
is possible to estimate
the interstellar radiation field from the equilibrium
dust temperature. We
adopt $T_\mathrm{ d}([140/90])$ as the equilibrium grain
temperature. The 140 $\mu$m--100 $\mu$m color of the
Galactic disk is concentrated around 2.0
\citep{hibi06,hirashita07}, which
corresponds to the dust temperature of 20.2 K for
$\beta =1$ and 17.3 K for $\beta =2$.
Assuming that the dust heating is dominated by
UV radiation \citep{buat96}, we estimate the UV
radiation field from the dust temperature.
The UV radiation field normalized to the
Galactic value is denoted as $G_\mathrm{ UV}$. Using the
scaling relation of
$G_\mathrm{ UV}\propto T_\mathrm{ d}^{4+\beta}$ \citep{evans95},
we obtain
\begin{eqnarray}
G_\mathrm{ UV}(\beta )=\left\{
\begin{array}{ll}
\left({\displaystyle \frac{T_\mathrm{ d}}{20.2~\mathrm{ K}}}
\right)^5 & \mbox{if $\beta =1$, } \\
\left({\displaystyle \frac{T_\mathrm{ d}}{17.3~\mathrm{ K}}}
\right)^6 & \mbox{if $\beta =2$.} \\
\end{array}
\right.
\end{eqnarray}

In Table \ref{tab:derived}, we present $G_\mathrm{ UV}(\beta )$
for $\beta =1$ and 2. Naturally we confirm the above conclusion
that the UV interstellar radiation field in BCDs is
systematically higher than that in spiral galaxies, although
Mrk 7 and II Zw 71 are exceptions. Moreover, there is little
difference in $G_\mathrm{ UV}$ between $\beta =1$ and 2. Thus,
it is more useful
to adopt $G_\mathrm{ UV}$ instead of $T_\mathrm{ d}$ to avoid
the dependence on $\beta$.

\begin{table*}
  \caption{Summary of the properties.}
  \label{tab:derived}
  \begin{center}
    \begin{tabular}{lcccccccc}
      \hline
      Name     & $D$ & \multicolumn{2}{c}{$G_\mathrm{ UV}$}
               & $L_\mathrm{ AKARI}$
               & \multicolumn{2}{c}{$L_\mathrm{ FIR}$} &
               \multicolumn{2}{c}{$M_\mathrm{ d}$} \\
               & & ($\beta =1$) & ($\beta =2$) & & ($\beta =1$) &
               ($\beta =2$) & ($\beta =1$) & ($\beta =2$) \\
               & [Mpc] & & & [$L_\odot$] & [$L_\odot$] & [$L_\odot$] &
               [$M_\odot$] & [$M_\odot$]
                              \\ \hline
      II Zw 40 & 9.2 &
               $120^{+400}_{-70}$ & $81^{+170}_{-42}$ & $6.2\times 10^8$
               & $1.1\times 10^9$
               & $8.1\times 10^8$ & $3.9\times 10^4$ & $1.1\times 10^5$ \\
      Mrk 7   & 42.4 &
              $5.4^{+4.9}_{-2.0}$ & $5.7^{+4.8}_{-2.0}$ & $1.9\times 10^9$
              & $2.3\times 10^9$
              & $2.2\times 10^9$ & $1.4\times 10^6$ & $3.7\times 10^6$ \\
      Mrk 71  & 3.4 &
              $39^{+77}_{-20}$ & $34^{+46}_{-16}$ & $4.4\times 10^7$ &
              $5.7\times 10^7$
              & $5.0\times 10^7$ & $5.4\times 10^3$ & $1.5\times 10^4$ \\
      UM 439  & 13.1 &
              $28^{+52}_{-14}$ & $25^{+35}_{-11}$ & $8.1\times 10^7$ &
              $1.0\times 10^8$
              & $9.1\times 10^7$ & $1.4\times 10^4$ & $3.7\times 10^4$ \\
      UM 533  & 10.9 &
              $17^{+29}_{-8}$ & $16^{+23}_{-7}$ & $8.1\times 10^7$ &
              $9.9\times 10^7$
              & $9.1\times 10^7$ & $2.2\times 10^4$ & $5.8\times 10^4$ \\
      II Zw 70 & 17.0 &
               $14^{+70}_{-8}$ & $14^{+49}_{-8}$ & $2.9\times 10^8$ &
               $3.4\times 10^8$
               & $3.2\times 10^8$ & $8.1\times 10^4$ & $2.1\times 10^5$ \\
      II Zw 71 & 18.5 &
              $1.4^{+0.9}_{-0.4}$ & $1.5^{+1.0}_{-0.5}$ & $2.5\times 10^8$
              & $3.9\times 10^8$
              & $3.5\times 10^8$ & $1.0\times 10^6$ & $2.5\times 10^6$ \\
      Mrk 36  & 8.2 &
              $>7.6$ & $>7.9$ & $2.2\times 10^7$ & ---\footnotemark[$*$]
              & ---\footnotemark[$*$]
              & $>8.2\times 10^2$ & $>1.5\times 10^3$ \\
      \hline
      NGC 2841 & 8.1 &
               $2.3^{+4.6}_{-1.1}$ & $2.5^{+4.6}_{-1.2}$ & $1.5\times 10^9$
               & $2.1\times 10^9$
               & $1.9\times 10^9$ & $3.7\times 10^6$ & $9.3\times 10^6$ \\
      NGC 2976 & 3.6 &
               $7.8^{+27}_{-4.3}$ & $8.1^{+23}_{-4.3}$ & $4.7\times 10^8$ &
               $5.8\times 10^8$
               & $5.5\times 10^8$ & $3.1\times 10^5$ & $8.1\times 10^5$ \\
      M 101    & 7.4 &
               $1.8^{+3.3}_{-0.8}$ & $2.0^{+3.4}_{-0.9}$ & $1.1\times 10^{10}$
               & $1.6\times 10^{10}$
               & $1.5\times 10^{10}$ & $3.4\times 10^7$ & $8.5\times 10^7$ \\
      \hline
       \multicolumn{9}{@{}l@{}}{\hbox to 0pt{\parbox{160mm}{\footnotesize
       Note. The last three galaxies are spiral galaxies for comparison
             (same as Table \ref{tab:Tdust}).
             Lower limits are shown by ``$>$''.
       \par\noindent
       \footnotemark[$*$] For Mrk 36, it is not possible to
             estimate $L_\mathrm{ FIR}$ because $T_\mathrm{ d}[140/90]$ is not
             determined.
    }\hss}}
    \end{tabular}
  \end{center}
\end{table*}

\subsection{Total FIR luminosity}\label{subsec:LFIR}

Total FIR luminosity reflects how much stellar light is absorbed
and reemitted by dust grains. It is empirically known that
FIR luminosity is a good indicator of SFR in
galaxies \citep{kennicutt98,inoue00}. Here we estimate the
total FIR luminosity of each BCD.
The FIS \textit{N60} ($65~\mu$m), \textit{WIDE-S}
($140~\mu$m),
and \textit{WIDE-L} ($140~\mu$m) bands have the
advantage of covering \textit{continuously} the
wavelength range from 50 $\mu$m to 170 $\mu$m
\citep{kawada07}.
Thus, by using those three bands,
we define the \textit{AKARI FIR flux}, $f_\mathrm{ AKARI}$, as an
estimate of the flux
in the above wavelength range:
\begin{eqnarray}
f_\mathrm{ AKARI} & \equiv & f_\nu (65~\mu\mathrm{ m})
\Delta\nu (\mbox{\textit{N60}})\nonumber\\
& & +
f_\nu (90~\mu\mathrm{ m})\Delta\nu (\mbox{\textit{WIDE-S}})
\nonumber\\
& & +
f_\nu (140~\mu\mathrm{ m})\Delta\nu (\mbox{\textit{WIDE-L}})\, ,
\end{eqnarray}
where $f_\nu (\lambda )$ is the flux per unit frequency at
wavelength $\lambda$, and $\Delta\nu (\mbox{band})$ denotes
the frequency width covered by the band. According to
\citet{kawada07}, we adopt
$\Delta\nu (\mbox{\textit{N60}})=1.58$ THz,
$\Delta\nu (\mbox{\textit{WIDE-S}})=1.47$ THz, and
$\Delta\nu (\mbox{\textit{WIDE-L}})=0.831$ THz. With
$f_\mathrm{ AKARI}$, we define the \textit{AKARI FIR luminosity},
$L_\mathrm{ AKARI}$, as an estimate of
the luminosity in $\lambda =50$--170 $\mu$m:
\begin{eqnarray}
L_\mathrm{ AKARI}\equiv 4\pi D^2f_\mathrm{ AKARI}\, ,
\end{eqnarray}
where $D$ is the distance to the object.
In Table \ref{tab:derived}, we list the estimated AKARI
FIR luminosities as well as the distances.
The distances are taken from
\citet{tolstoy95} for Mrk 71 (NGC 2366),
\citet{osman82} for NGC 2841,
\citet{karachentsev02} for NGC 2976, and
\citet{jurcevic06} for M 101. For more distant
galaxies, we estimate the distance from the Galactocentric
velocity (taken from
NED)\footnote{http://nedwww.ipac.caltech.edu/.} by
assuming a Hubble constant of
75 km s$^{-1}$ Mpc$^{-1}$.

It is also possible to estimate the FIR luminosity including
$\lambda >170~\mu$m by assuming a functional form of
$\propto\nu^\beta B_\nu (T_\mathrm{ d})$, where we adopt
$T_\mathrm{ d}=T_\mathrm{ d}([140/90])$. Then we obtain an estimate
of the total FIR luminosity at $\lambda >50~\mu$m. Following
the procedure in
\citet{nagata02}, we estimate the total FIR luminosity as
below.

First, we estimate the total FIR flux. The total FIR flux,
$f_\mathrm{ FIR}$, can be well approximated by \citep{nagata02}
\begin{eqnarray}
f_\mathrm{ FIR} & \simeq & \int_0^\infty\Omega\tau_{90}\left(
\frac{\nu}{\nu_{90}}\right)^\beta B_\nu(T_\mathrm{ d})\, d\nu
\nonumber\\
& + & \Biggl[f_\nu (65~\mu\mathrm{ m})-B_\nu (65~\mu\mathrm{ m},\,
T_\mathrm{ d})\Biggr.\nonumber\\
& & \Biggl.\times\Omega\tau_{90}\left(\frac{65}{90}
\right)^{-\beta}
\Biggr]\Delta\nu (\mbox{\textit{N60}})\, ,
\end{eqnarray}
where $\Omega$ is the solid angle of the object, $\tau_{90}$
is the optical depth at $\lambda =90~\mu$m, $\nu_{90}$ is the
frequency corresponding to $\lambda =90~\mu$m
(i.e., 3.33 THz), and
$B_\nu (\lambda ,\, T_\mathrm{ d})$ is the Planck function
evaluated at wavelength $\lambda$.
The first term represents the total FIR flux from
grains in radiative equilibrium, and the second term shows
the contribution of the excess emission of stochastically
heated grains in \textit{N60} (65 $\mu$m). Thus, $f_\mathrm{FIR}$
provides the total FIR flux for $\lambda >50~\mu$m.
Noting that
$f_{\nu}(90~\mu\mathrm{ m})=\Omega\tau_{90}B_\nu
(90~\mu\mathrm{ m},\, T_\mathrm{ d})$, we obtain the following
expression similar to that of \citet{nagata02}:
\begin{eqnarray}
f_\mathrm{ FIR} & \simeq & C_1f_\nu (65~\mu\mathrm{ m})
\nonumber\\
& + & f_\nu (90~\mu\mathrm{ m})\left[\exp\left(
\frac{159.8}{T_\mathrm{ d}}\right)-1\right]\nonumber\\
& & \times\left[C_2T_\mathrm{ d}^{\beta +4}-
\frac{C_3}{\exp (221.3/T_\mathrm{ d})-1}\right]\, ,
\end{eqnarray}
where $C_1=1.58\times 10^{12}$ Hz,
$C_2=795.6~\mathrm{ Hz}~\mathrm{ K}^{-5}$ and
24.42~Hz~K$^{-6}$ for $\beta =1$ and 2, respectively, and
$C_3=8.262\times 10^{12}$ Hz and $1.144\times 10^{13}$ Hz
for $\beta =1$ and 2, respectively.

By using $f_\mathrm{ FIR}$, we obtain the total FIR luminosity
$L_\mathrm{ FIR}$ as
\begin{eqnarray}
L_\mathrm{ FIR}=4\pi D^2f_\mathrm{ FIR}\, ,
\end{eqnarray}
where $D$ is the distance to the object. In
Table \ref{tab:derived}, we present the estimated FIR
luminosities for the sample. The FIR luminosities estimated
with $\beta =1$ and 2 are indicated by
$L_\mathrm{FIR}(\beta =1)$ and $L_\mathrm{FIR}(\beta =2)$,
respectively. For Mrk 36, it is not possible to determine
$L_\mathrm{FIR}$ since $T_\mathrm{d}$ cannot be determined.
Thus, we exclude Mrk 36 from the analysis here. The error
of $L_\mathrm{FIR}$ mainly comes from the uncertainty in the
flux estimation
($\sim 20$\%). We note that the uncertainty caused by
the difference in $\beta$ is comparable to or smaller than
20\%.

It is convenient to obtain a conversion formula from the
AKARI FIR luminosity to the total FIR luminosity. We define
the ``correction'' factor, $C_\mathrm{ AKARI}$, which is to be
multiplied to
the AKARI FIR luminosity to obtain the total FIR luminosity
as
\begin{eqnarray}
L_\mathrm{ FIR}(\beta )=C_\mathrm{ AKARI}(\beta )L_\mathrm{ AKARI}\, ,
\end{eqnarray}
where $L_\mathrm{ FIR}$ and thus $C_\mathrm{ AKARI}$ depend on the
assumed value of $\beta$. Because the sample size is
small, we cannot investigate the possible dependence of
$C_\mathrm{ AKARI}$ on various
physical parameters such as dust temperature. Thus,
we only estimate the average of $C_\mathrm{ AKARI}$ for
the sample. The results are listed in
Table \ref{tab:cakari}, where the averages of
$C_\mathrm{ AKARI}$ with and
without the spiral sample are shown. We see that there is no
significant difference in $C_\mathrm{ AKARI}$ between those
two averages. On the other hand, there seems to be
difference in
$C_\mathrm{ AKARI}$ between $\beta =1$ and 2, mainly because
the extrapolation to long wavelengths
($\lambda >170~\mu$m) in estimating the total FIR luminosity
depends on $\beta$. In summary, we can convert the AKARI
FIR luminosity to the total FIR luminosity by multiplying
1.2--1.4. We note that a further 30--40\%
correction for
the contribution from mid-infrared emission at
$\lambda <50~\mu$m will be necessary
to estimate the total infrared luminosity of dust
emission \citep{dale02}.

\begin{table}
  \caption{Correction factor ($C_\mathrm{ AKARI}$) multiplied to
the AKARI FIR luminosity
to obtain the total FIR luminosity.}
  \label{tab:cakari}
  \begin{center}
    \begin{tabular}{c|cc}
      \hline
           & \multicolumn{2}{@{}c@{}}{Spiral sample} \\
      $\beta$ & included & not included \\ \hline
      1  & $1.36\pm 0.17$ & $1.34\pm 0.19$ \\
      2  & $1.19\pm 0.10$ & $1.22\pm 0.11$ \\
      \hline
       \multicolumn{3}{@{}l@{}}{\hbox to 0pt{\parbox{65mm}{\footnotesize
       Note. The number after ``$\pm$'' shows the standard deviation
             (1 $\sigma$).
    }\hss}}
    \end{tabular}
  \end{center}
\end{table}

\subsection{Star formation rate}\label{subsec:sfr}

It is empirically known that the FIR luminosity is a good
indicator of the SFR. \citet{hirashita03} provide a
conversion formula from $L_\mathrm{ FIR}$ to SFR (we assume
that $L_\mathrm{ IR}$ in
\cite{hirashita03} is approximated by $L_\mathrm{ FIR}$,
but as mentioned in section \ref{subsec:LFIR}, the upward
30--40\% correction for $\lambda <50~\mu$m may be
required for $L_{\rm FIR}$):
\begin{eqnarray}
\mathrm{ SFR}=C_\mathrm{ IR}L_\mathrm{ FIR}\, ,
\label{eq:sfr}
\end{eqnarray}
where we adopt the conversion coefficient
$C_\mathrm{IR}=2.0\times 10^{-10}~M_\odot~\mathrm{yr}^{-1}~
L_\odot^{-1}$ as recommended by \citet{hirashita03}.
The SFR estimated by using $L_\mathrm{FIR}(\beta =1)$ is
listed in Table \ref{tab:discussion} for each galaxy.
If we adopt $L_\mathrm{FIR}(\beta =2)$, we obtain smaller
SFR by 4--26\%. The range of the SFR is
0.01--0.4 $M_\odot$ yr$^{-1}$ for the present BCD
sample. This is within the range (a few times $10^{-3}$
to several times 10 $M_\odot$ yr$^{-1}$) given for a BCD
sample by \citet{hopkins02}.

\begin{table*}
  \caption{Some related quantities used for discussion.}
  \label{tab:discussion}
  \begin{center}
    \begin{tabular}{lcccccc}
      \hline
      Name     & SFR\footnotemark[$*$] & $M_{\mathrm{H}\emissiontype{I}}$
               & $\tau_\mathrm{ SF}$
               & $12+\log\mathrm{ (O/H)}$ & ${\cal D}$ \\
               & [$M_\odot$ yr$^{-1}$] & [$M_\odot$] & [Gyr] & &
                              \\ \hline
      II Zw 40 & 0.22 &
               $2.0\times 10^8$ & 0.91 & 8.15
               & $2.0\times 10^{-4}$ \\
      Mrk 7   & 0.50 &
              $3.6\times 10^9$ & 7.2 & 8.54
              & $3.9\times 10^{-4}$ \\
      Mrk 71  & 0.012 &
              $1.2\times 10^9$ & 110 & 7.83
              & $4.5\times 10^{-6}$ \\
      UM 439  & 0.021 &
              $1.7\times 10^8$ & 8.1 & 7.98
              & $8.2\times 10^{-5}$ \\
      UM 533  & 0.020 &
              $5.8\times 10^7$ & 2.9 & 8.10
              & $3.8\times 10^{-4}$ \\
      II Zw 70 & 0.073 &
              $3.6\times 10^8$ & 4.9 & 8.11
              & $2.3\times 10^{-4}$ \\
      II Zw 71 & 0.080 &
              $7.3\times 10^8$ & 9.1 & 8.24
              & $1.4\times 10^{-3}$ \\
      Mrk 36  & ---\footnotemark[$\dagger$] &
              $1.8\times 10^7$ & $<2.5$ & 7.81 & $>4.6\times 10^{-5}$
              \\
      \hline
       \multicolumn{6}{@{}l@{}}{\hbox to 0pt{\parbox{120mm}{\footnotesize
       Note. Lower and upper limits are shown by ``$>$'' and ``$<$'',
respectively. For Mrk 36, SFR cannot be estimated.
       \par\noindent
       \footnotemark[$*$] The SFR is derived by using
$L_\mathrm{ FIR}(\beta =1).$ If $L_\mathrm{ FIR}(\beta =2)$ is used instead,
the SFR becomes 4--26\% higher according to the difference in
$L_\mathrm{ FIR}$.
       \par\noindent
       \footnotemark[$\dagger$] The SFR of Mrk 36 cannot be estimated
because $L_\mathrm{ FIR}$ is not available.
    }\hss}}
    \end{tabular}
  \end{center}
\end{table*}

There are two common BCDs between this paper and
\citet{hopkins02}: Mrk 71 and
Mrk 36. For the latter galaxy, we do not
obtain the SFR because it is not possible to
determine $L_\mathrm{ FIR}$ (section \ref{subsec:LFIR}).
For Mrk 71, our estimate is
2--3 times lower than their estimate
(0.065 $M_\odot$ yr$^{-1}$ from the
IRAS 60 $\mu$m luminosity and
0.041 $M_\odot$ yr$^{-1}$ from the 1.4 GHz
radio luminosity).
This confirms the comment of
\citet{hirashita03}
that there is a risk of underestimating the SFR for
$\mathrm{ SFR}\ltsim 1~M_\odot$ yr$^{-1}$ if
we use equation (\ref{eq:sfr}). This may be
because of smaller dust optical depth.

There are three overlapping BCDs between this paper
and \citet{sage92}: II Zw 40, UM 439, and UM 533.
Their SFR derived from the FIR luminosity (0.46,
0.085, and 0.049 $M_\odot$ yr$^{-1}$, respectively)
is systematically higher, which
is attributed to different $C_\mathrm{ IR}$
($6.5\times 10^{-9}~M_\odot~\mathrm{ yr}^{-1}~L_\odot^{-1}$).
This high $C_\mathrm{ IR}$ is consistent with the
SFR derived from H$\alpha$ luminosity in
dwarf galaxies \citep{thronson86}, which again confirms
the necessity of using higher $C_\mathrm{ IR}$ than
proposed by \citet{hirashita03} for
$\mathrm{ SFR}\ltsim 1~M_\odot$ yr$^{-1}$ .
Moreover, the SFR derived from H$\alpha$ in
\citet{sage92} (3.1, 0.20, and 0.19 $M_\odot$ yr$^{-1}$,
respectively) is much higher than SFR derived from the
FIR luminosity, which is again explained by
small dust optical depth as mentioned in the previous
paragraph. However, the SFR derived
from H$\alpha$ luminosity in \citet{sage92} could be
overestimated because they assume that 2/3 of
the ionizing photons are absorbed by dust, which might
be less in low-metallicity galaxies like BCDs.

\subsection{Dust mass}\label{subsec:Mdust}

As stated in section \ref{sec:intro}, the
\textit{WIDE-S} ($90~\mu$m) and \textit{WIDE-L}
($140~\mu$m) bands are suitable to trace the
total mass of large grains which have the dominant
contribution to the total dust content.
The total dust mass $M_\mathrm{d}$ is related to the
flux as (e.g., \cite{hildebrand83})
\begin{eqnarray}
M_\mathrm{ d}=
\frac{F_\nu (\lambda )D^2}{\kappa_\nu B_\nu (T_\mathrm{ d})}
\, ,
\end{eqnarray}
where $\kappa_\nu$ is the mass absorption coefficient of
dust grains.
For simplicity, we assume spherical grains with a single
radius $a$ and with a uniform density $s$.
As shown later, the assumption on $a$ is not essential,
since $\kappa_\nu$ is insensitive to $a$.
The cross section of a grain is expressed as
$\pi a^2Q_\nu$, where $Q_\nu$ is a dimensionless
quantity expressing the emission efficiency.
Using $Q_\nu$, the
mass absorption coefficient is written as
\begin{eqnarray}
\kappa_\nu =\frac{3Q_\nu}{4as}\, .
\end{eqnarray}
Since there is a relation of $Q_\nu\propto a$ for
$a\ll\lambda$
\citep{evans95}, the estimate of $\kappa_\nu$ is insensitive
to the assumed value of $a$.
\citet{hildebrand83} estimates $Q_\nu /a$ at $\lambda =125~\mu$m
to be 3/400 $\mu$m$^{-1}$. We note that the normalization of
$Q_\nu /a$ at $\lambda =125~\mu$m is convenient since it is
roughly the middle of the two wavelengths ($\lambda =90~\mu$m
and 140 $\mu$m) used for the estimate of dust mass. If we adopt
$s=3$ g cm$^{-3}$, we obtain
\begin{eqnarray}
\kappa_\nu =18.8\left(\frac{\lambda}{125~\mu\mathrm{ m}}
\right)^{-\beta}~\mathrm{ cm}^{2}~\mathrm{ g}^{-1}\, .
\end{eqnarray}
In Table \ref{tab:derived}, we list the dust mass,
where we adopt $T_\mathrm{ d}([140/90])$ for the dust temperature.
The dust mass estimated with $\beta =1$ and 2 are denoted
as $M_\mathrm{ d}(\beta =1)$ and $M_\mathrm{ d}(\beta =2)$,
respectively.
The dust mass in Mrk 36 is given by using the
$T_\mathrm{ d}([65/90])$ instead of $T_\mathrm{ d}([140/90])$,
for which only a lower limit is given. Since
$T_\mathrm{ d}([65/90])$ tends to overestimate the
equilibrium grain temperature (section \ref{subsec:Tdust}),
we regard this dust mass to be a lower limit.
{}From Table \ref{tab:derived}, we observe that
$M_\mathrm{ d}(\beta =1)$ is systematically lower than
$M_\mathrm{ d}(\beta =2)$ by a factor of 2.5.
This is because higher dust temperature for $\beta =1$
leads to a lower dust mass with a fixed dust emissivity.

One of our BCDs, II Zw 40, overlaps with the sample in
\citet{galliano05} (section \ref{subsec:remark}), who model
the dust SED. First of all,
the mass of the big grain component, which is contributing
to the FIR regime, is just the same as our estimate
($1.1\times 10^5M_\odot$) if we adopt $\beta =2$.
This means that the AKARI \textit{WIDE-S} ($90~\mu$m)
and \textit{WIDE-L} ($140~\mu$m) bands are indeed
useful to trace the dust mass.
However, they also show an additional very cold dust
component contributing
to the emission at submillimeter wavelengths, although
the amount of this component is dependent on the FIR emissivity
index $\beta$. This very
cold component cannot be investigated by AKARI, and
we concentrate on the dust component contributing to the
FIR emission in this paper.

\section{Discussion}\label{sec:discussion}

\subsection{Star formation properties}
\label{subsec:consumption}

In section \ref{subsec:ISRF}, we have shown that the BCD
sample tends to show higher UV radiation fields than spiral
galaxies. This is consistent with a picture that intense star
formation is occurring in a concentrated region in BCDs.
In order to examine the properties of such an intense
star formation activity (e.g.,
a burst or a continuous mode of star formation)
it is useful to estimate the gas consumption timescale
($\tau_\mathrm{ SF}$) as follows.

Here, we adopt the SFR estimated in
section \ref{subsec:sfr} by adopting
$L_\mathrm{ FIR}(\beta =1)$. (If we adopt
$L_\mathrm{ FIR}(\beta =2)$, the gas consumption timescale
becomes 4--26\% shorter.) Then the
gas consumption timescale is defined as the gas mass
divided by the SFR. For the gas mass, we adopt
the mass of H\emissiontype{I} gas
($M_{\mathrm{H}\emissiontype{I}}$) in the same way
as in \citet{hirashita02} and \citet{hopkins02},
since molecular gas is rarely detected for BCDs
\citep{barone00}.
We estimate $M_{\mathrm{H}\emissiontype{I}}$ of II Zw 71
by using the
H\emissiontype{I}
flux measured by \citet{huchtmeier89} and by using
equation (1) in \citet{lisenfeld98}.
The estimated gas consumption time $\tau_\mathrm{ SF}$
is listed in Table \ref{tab:discussion}.
We observe that $\tau_\mathrm{ SF}$ is distributed with a
range of two orders of magnitude, which is consistent with
\citet{hopkins02}. Moreover, the gas consumption time does not
correlate with $G_\mathrm{ UV}$. For example, Mrk 71 has
large $G_\mathrm{ UV}$, which indicates a concentrated intense
star formation, but has large $\tau_\mathrm{ SF}$ exceeding the
cosmic timescale. Such galaxies with long $\tau_\mathrm{ SF}$
but high $G_\mathrm{ UV}$ give us a picture that the
star formation is strongly activated only in a limited region.
On the contrary, II Zw 40 has large $G_\mathrm{ UV}$ and small
$\tau_\mathrm{ SF}$, indicating the whole
galaxy may be in an active phase of star formation.

It is possible that the star formation in BCDs is
intermittent. In this case, the diversity in
$\tau_\mathrm{SF}$ is naturally
explained: if a BCD is in activated/inactivated phase of
star formation, $\tau_\mathrm{SF}$ becomes small/large.
Indeed, the effect of feedback is considered to be enhanced
in small systems such as BCDs, leading to an intermittent
nature of star
formation \citep{saito00,carraro01,kobayashi04,kamaya05}.
It is also observationally suggested that the star formation
in dwarf galaxies may be episodic
\citep{fanelli88,greggio98,grossi07}.

\subsection{Dust-to-gas ratio and chemical enrichment}
\label{subsec:dust_metal}

The dust production is strongly related to the metal
production. Indeed, the relation between
dust abundance (usually dust-to-gas ratio is adopted)
and metal abundance (metallicity) can be
investigated by
using chemical evolution models \citep{dwek98,hirashita02}.
Here we consider whether dust-to-gas ratio and
metallicity are related or not.

As the metal abundance, we adopt oxygen abundance, which
is the most easily observed. The oxygen abundance of each
sample BCD is compiled in
\citet{hirashita02}
and \citet{hopkins02}, but for II Zw 71 we take the
oxygen abundance from \citet{kewley05}. The dust-to-gas ratio
${\cal D}$ is estimated by $M_\mathrm{d}(\beta =1)$ divided by
$M_{\mathrm{H}\emissiontype{I}}$. If we use
$M_\mathrm{d}(\beta =2)$ instead,
we obtain about 2.5 times higher dust-to-gas ratio.
Note that the dust-to-gas ratio would shift further upward
if there is really a very cold dust component contributing
to longer wavelengths than those traced by AKARI
(\cite{galliano05}; section \ref{subsec:Mdust}).
In Table \ref{tab:discussion}, we list the metallicity
(oxygen abundance: $12+\log (\mathrm{O/H})$) and the
dust-to-gas ratio ${\cal D}$. Note that the solar abundance is
$12+\log (\mathrm{O/H})=8.93$ \citep{anders89}.

\begin{figure}
  \begin{center}
    \FigureFile(85mm,85mm){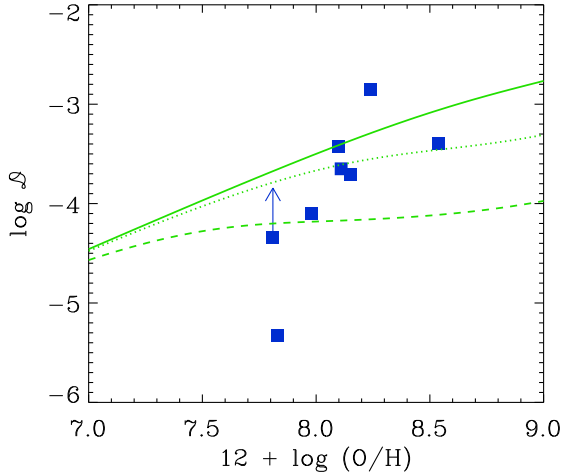}
  \end{center}
  \caption{Relation between dust-to-gas ratio ${\cal D}$
and metallicity. The data of the present BCD sample are
shown by solid squares. The arrow indicates a lower limit.
The solid, dotted, and dashed lines show the model
predictions by \citet{hirashita02}
with various dust destruction efficiencies
(in their notation $\beta_\mathrm{ SN}=1$, 5, and 25,
respectively).}
  \label{fig:dustmetal}
\end{figure}

In Figure \ref{fig:dustmetal}, we show the relation between
those two quantities. We observe that there is a positive
correlation between dust-to-gas ratio and metallicity. This
correlation is
also shown by \citet{engelbracht08} with a larger metallicity
range. The correlation is consistent
with the picture that dust enrichment proceeds as
the system is enriched by metals.

It is also interesting to compare the results with theoretical
model predictions. Here we adopt a model by
\citet{hirashita02} (originally developed by \cite{lisenfeld98}),
who consider a one-zone chemical evolution model with
the instantaneous recycling approximation \citep{tinsley80}.
The model equations consist of the evolutions of 
gas mass, metal mass, and dust mass. By treating these
equations, the relation between dust-to-gas ratio
and metallicity is obtained. In particular, there are two
important parameters in the model: (i) the fraction of
metals condensed into dust grains, $f_\mathrm{in}$, and
(ii) the ``dust destruction efficiency'', $\beta_\mathrm{SN}$.
The former parameter is fixed to 0.1, and does not contribute
to the scatter of the dust-to-gas ratio if the condensation
efficiency in the stellar mass loss is independent of age.
The ``dust destruction efficiency'' $\beta_\mathrm{ SN}$ is
defined as
$\beta_\mathrm{ SN}\equiv M_\mathrm{ g}/(\tau_\mathrm{ SN}\psi)$,
where $M_\mathrm{ g}$ is the total gas mass, $\tau_\mathrm{ SN}$
is the timescale of dust destruction by supernova
shocks, and $\psi$ is the SFR. This parameter can be
expressed as follows \citep[equation 6]{hirashita02}:
\begin{eqnarray}
\beta_\mathrm{SN}=\epsilon M_\mathrm{s}
\frac{\gamma}{\psi}\frac{1}{X_{\rm SF}}\, ,
\end{eqnarray}
where $\epsilon$ is the fraction of destroyed dust in a SN
blast, $M_{\rm s}$ is the gas mass accelerated to a velocity
large enough for dust destruction (typically
100 km s$^{-1}$), $\gamma$ is the supernova rate, and
$X_{\rm SF}$ is the gas mass fraction in the star-forming
region, where the dust formation and destruction are
really occurring (the rest of the gas is considered to be
contained in the H \textsc{i} envelope, which they assume not
to be concerned with the
dust formation and destruction). Typical values are
$\epsilon\sim 0.1$ \citep{mckee89},
$M_{\rm s}\sim 6800M_\odot$ (if we adopt $10^{51}$ erg
for the kinetic energy input of a supernova, and
100 km s$^{-1}$ for the threshold velocity for the dust
destruction), $\gamma/\psi\sim 1/136M_\odot$ (for
the Galaxy; \cite{lisenfeld98}), we obtain
$\beta_\mathrm{SN}=5$ for $X_{\rm SF}=1$.
\citet{hirashita02}
argue that this parameter changes according to the
relative contribution between Type Ia
supernovae and Type II supernovae. If an intermittent
star formation activity is assumed, $\gamma/\psi$
can change by a factor of 20
\citep{bradamante98,hirashita02}. Motivated by this
factor, they changed
$\beta_\mathrm{ SN}$ from 1 to 25.

In Figure \ref{fig:dustmetal}, we show the theoretical
predictions for $\beta =1$, 5, and 25 with $f_{\rm in}=0.1$.
We observe that those models explains the data points except
for the two extremes (II~Zw~71 and Mrk~71 for the upper
and lower extremes in Figure \ref{fig:dustmetal}, respectively).
The variation in
$\beta_\mathrm{ SN}$ increases the scatter of the dust-to-gas
ratio at a certain metallicity and makes the correlation less
clear. Further change of $\beta_{\rm SN}$ may be possible because
of the change of $X_{\rm SF}$. As mentioned in
section \ref{subsec:consumption}, only a small fraction of gas
may be
concerned with star formation in Mrk 71. Thus, $X_{\rm SF}\ll 1$ is
expected for Mrk 71, where
$\beta_{\rm SN}$ could be larger than the above values.
Such a large value of $\beta_{\rm SN}$ may explain the
extremely poor dust-to-gas ratio of Mrk 71.

As commented above, the dust-to-gas ratios are larger if we adopt
$\beta =2$, and/or if the cold dust component contributing to
the submillimeter wavelengths, which we fail to trace in FIR,
really exists as proposed by \citet{galliano05}.
In these cases, we should adopt a larger $f_{\rm in}$, but a
large variation of $\beta_{\rm SN}$ is still required to explain
the large scatter of dust-to-gas ratios.

\section{Summary}\label{sec:summary}

We have reported far-infrared (FIR) properties of eight blue
compact dwarf galaxies (BCDs) observed by AKARI.
The fluxes at
wavelengths of 65 $\mu$m, 90 $\mu$m, 140 $\mu$m, and
160 $\mu$m are measured and are used to
estimate basic quantities about dust
grains such as dust temperature, dust mass,
and total FIR luminosity. We have found that the typical dust
temperature of our BCD sample, except for
Mrk 7 and II Zw 71, is significantly higher than
that of normal spiral galaxies. This indicates high
interstellar radiation field, which proves to range up to 100
times the Galactic value.
This confirms the concentrated star-forming activity in
BCDs. The star formation rate (SFR) is also evaluated from the FIR
luminosity as 0.01--0.5 $M_\odot$ yr$^{-1}$. Combining this
quantity with
gas mass taken from the literature, we estimate the gas consumption
timescales (gas mass divided by SFR), which
span a wide range from
1 Gyr to 100 Gyr. A natural
interpretation of this large variety can be provided by
intermittent star formation activity. We finally show the relation
between dust-to-gas ratio
and metallicity (we utilize our estimate of dust mass, and take
other necessary quantities from the literature). There is a
positive correlation between dust-to-gas ratio and
metallicity as expected from chemical evolution models.

\vspace{3mm}

We are grateful to S. Madden, the referee, for her
helpful comments that improved this paper very much.
We thank H. Shibai, Y. Y. Tajiri, M. Nagaoka, Y.~Doi,
S. Matsuura, and M. Shirahata for useful discussions about
observations and
data analysis. We thank all members of AKARI project for
their continuous help and support. This research has made
use of the NASA/IPAC Extragalactic Database (NED), which
is operated by the Jet Propulsion Laboratory, California
Institute of Technology, under contract with the National
Aeronautics and Space Administration.




\end{document}